\documentclass[lettersize,journal]{IEEEtran}
\usepackage{amsmath,amsfonts}
\usepackage{algorithmic}
\usepackage{array}
\usepackage[caption=false,font=normalsize,labelfont=sf,textfont=sf]{subfig}
\usepackage{textcomp}
\usepackage{stfloats}
\usepackage{url}
\usepackage{verbatim}
\usepackage{graphicx}
\def\BibTeX{{\rm B\kern-.05em{\sc i\kern-.025em b}\kern-.08em
    T\kern-.1667em\lower.7ex\hbox{E}\kern-.125emX}}
\usepackage{balance}
\usepackage{booktabs}
\usepackage[colorlinks, linkcolor=blue, citecolor=blue, urlcolor=blue]{hyperref}
\begin{document}

\title{Classification of Power Quality Disturbances Using Resnet with Channel Attention Mechanism}

\author{Su Pan, Xingyang Nie, Xiaoyu Zhai, Biao Wang, Huilin Ge, Cheng He and Zhenping Ding 

\thanks{Xingyang Nie, Su Pan, Biao Wang, Huilin Ge, and Cheng He are with the Ocean College, Jiangsu University of Science And Technology, Zhenjiang 212003, China. E-mail: starsun87@126.com, 1242425221@qq.com, wangbiao@just.edu.cn, ghl1989@just.edu.cn, hecyz@qq.com.

Xiaoyu Zhai is with Nanjing Research Institute of  Electronic Equipment,
China Aerospace Science and Industrial Corporation, Nanjing 210000, China. E-mail: 573911383@qq.com.

Zhenping Ding, Nanjing University of Science and Technology,  Zijin College, Nanjing 210023, China. E-mail:dzpseu@139.com}
\thanks{(Corresponding author: Xingyang Nie.)}}

\maketitle

\begin{abstract}
The detection and classification of power quality disturbances (PQDs) carries significant importance for power systems. In response to this imperative, numerous intelligent diagnostic methods have been developed. However, existing identification methods usually concentrate on single-type signals or on complex signals with two types, rendering them susceptible to noisy labels and environmental effects. This study proposes a novel method for the classification of PQDs, termed ST-GSResNet, which utilizes the S-Transform and an improved residual neural network (ResNet) with a channel attention mechanism. The ST-GSResNet approach initially uses the S-Transform to transform a time-series signal into a 2D time-frequency image for feature enhancement. Then, an improved ResNet model is introduced, which employs grouped convolution instead of the traditional convolution operation. This improvement aims to facilitate learning with a block-diagonal structured sparsity on the channel dimension, the highly-correlated filters are learned in a more structured way in the networks with filter groups.  By reducing the number of parameters in the network in this significant manner, the model becomes less prone to overfitting. Furthermore, the SE module concentrates on primary components, which enhances the model's robustness in recognition and immunity to noise. Experimental results demonstrate that, compared to existing deep learning models, our approach has advantages in computational efficiency and classification accuracy. 
\end{abstract}


\section{Introduction}
\IEEEPARstart{R}{apid}  industrialization continues to lead to an increase in greenhouse gas emissions, further intensifying global warming. The urgent task at hand is to implement a comprehensive global dual-carbon strategy to address this challenge. This strategy entails achieving carbon peaking and carbon neutrality, promoting the development and utilization of new energy sources, and enhancing energy efficiency and conservation. However, the integration of new intermittent renewable energy sources, electric vehicles, and energy storage devices into the grid can deteriorate power quality \cite{wang2019}; This poses substantial economic challenges and safety risks are posed for both power users and grid companies \cite{lin2014}. Therefore, a crucial step in addressing power quality issues involves accurately identifying power quality disturbances (PQDs).

Power quality is measured by deviations from defined standards in electrical parameters, including voltage, current, and electromagnetic fields, within the power system. PQDs encompass fluctuations in these electrical parameters, primarily induced by the operation of nonlinear loads, switching devices, and momentary faults. Such disturbances can negatively impact the reliability, production efficiency, and lifespan of equipment within the power system, as well as the safety of humans and property. The prevalence of renewable energy sources and power electronics devices in electricity grids has given rise to a "dual high" power system, which aims to enhance grid flexibility and cleanliness; however, it may have a detrimental effect on power quality by introducing high-frequency harmonics, spikes, and dips \cite{liu2021}. In complex grid environments, PQDs seldom occur in isolation. Instead, often combine fundamental PQDs, resulting in mixed power quality interference, also known as composite or mixed PQDs. 

Additionally, these disturbances often involve transient components. Analyzing and identifying mixed PQDs poses a significant challenge due to the intricate interplay between the characteristics of time and frequency domains. Effective management of PQDs is crucial from a practical standpoint \cite{xu2022}.

To address this issue, various signal processing techniques, such as the short-time Fourier transform (STFT) \cite{zhong2019}, wavelet transform (WT) \cite{chen2022}, Hilbert-Huang transform (HHT) \cite{li2016}, empirical mode decomposition (EMD) \cite{zhong2019}, and variational mode decomposition (VMD) \cite{zhao2019}, have been employed for parameter detection and feature analysis. Among these techniques, the S-Transform (ST) \cite{li2021} combines the strengths of STFT, WT, and Fourier transform (FT) \cite{larissa2020}, while overcoming certain limitations of the wavelet transform. Consequently, the ST transform has widespread application in identifying PQDs.In \cite{Liu2023}, a PQD classification scheme based on Fully Convolutional Networks (FCN) and Bidirectional Gated Recursive Units (BiGRU) is proposed. In \cite{uckol2023}, the author proposed a method based on deep learning and 2D wavelet scale images for classifying PQDs. The two aforementioned deep learning models employ distinct classification strategies: one directly utilizes a 1D sampling sequence, while the other transforms it into a 2D image using visualization techniques. Compared to the former, the latter strategy offers the advantage of potentially presenting a more comprehensive representation of the PQD signals' characteristics, such as spectral and nonlinear features, thereby enhancing the model's classification performance. Furthermore, the method of converting one-dimensional sequences into two-dimensional images provides superior interpretability, allowing for an intuitive understanding of the PQD signal characteristics through observation of the generated images.

The paper outlines several significant contributions: (i) A new method, ST-GSResNet, for the identification of power quality disturbance signals is proposed, which combines the S-Transform with an improved ResNet. This method demonstrates robustness to noise and maintains satisfactory performance even under conditions of noisy labels and environments. (ii) This paper introduces a novel approach that uses multi-resolution analysis with the S-Transform to encode one-dimensional time series signals into two-dimensional images. (iii) In this paper, we introduce group convolution to replace the original convolution, and incorporate the Squeeze and Excite (SE) module into the network, which significantly improves the model's identification and generalization ability. The proposed method aims to enhance the time-frequency characteristics of Power Quality Disturbance signals and effectively employ Deep Residual Networks (ResNet) to learn advanced features.

The remainder of this paper is structured as follows. Section 2 describes the background. Section 3 describes the research methodology and development. Section 4 shows the results and analysis. Finally, Section 5 concludes this paper.

\section{RELATED WORK}
This section introduces the theoretical topics that are essential for understanding the progress of this work. The discussion begins with the PQDs and extends to other topics pertinent to this research.
\subsection{S-Transform(ST)}

The S-Transform \cite{stockwell1996}, initially proposed by the American mathematician R.L. Bellman in the 1950s, is a mathematical method employed for signal analysis. It processes both continuous-time and discrete-time signals. The principle of the S-Transform involves converting a signal from the time domain (or spatial domain) to the frequency domain. In the frequency domain, signal characteristics are represented by amplitude and phase, facilitating a more thorough understanding and processing of the signal's frequency components. The S-Transform can be considered a spectral analysis method that combines the characteristics of the short-time Fourier transform and wavelets. It achieves multi-resolution signal analysis effectively, obtaining the accurate phase of each frequency component. By utilizing a window with width inversely proportional to frequency, the S-Transform can provide high time resolution for high-frequency components and high-frequency resolution for low-frequency signal components. Since most complex power quality events are non-stationary, the S-Transform effectively extracts features by employing an adaptable transformation with a Gaussian window.

The formula for the S-Transform is as follows:

\begin{equation}
\label{deqn_ex1}
S(a, b)=\int\left[f(t) g^*(t-b) e^{-2 \pi i a t}\right] d t
\end{equation}

The output obtained after applying the S-Transform is denoted as -S(a, b), where a and b are two parameters utilized in the S-Transform. -f(t) corresponds to the original signal. Moreover, -g(t) represents a Gaussian window with a width that inversely varies with frequency within the S-Transform. Lastly, -a  represents the frequency parameter.
The formula for the S-Transform illustrates its process of analyzing signals in the frequency domain. Adjusting the values of parameters a and b enables customization of the analysis window's frequency and time resolution to better match the characteristics of the signal. 
\subsection{ResNet}
ResNet \cite{he2016} is employed as a crucial variant of CNN for classifying images with a magnitude map, which describes voltage perturbation signals. Compared to traditional CNNs, ResNet incorporates residual connections to link the inputs and outputs of different layers, aiming to address the common issues known as gradient vanishing and Gradient explosion that can arise during training in many deep neural networks. In a Residual Network, the inputs and outputs of each stage are referred to as feature maps, generated by a series of convolutional layers, ReLU layers, and pooling layers. The output of each block represents a feature extraction of the input image, enabling the residual network to further increase its depth further and achieve superior performance.

\begin{figure}[ht]
\centering
\includegraphics[width=3.5in]{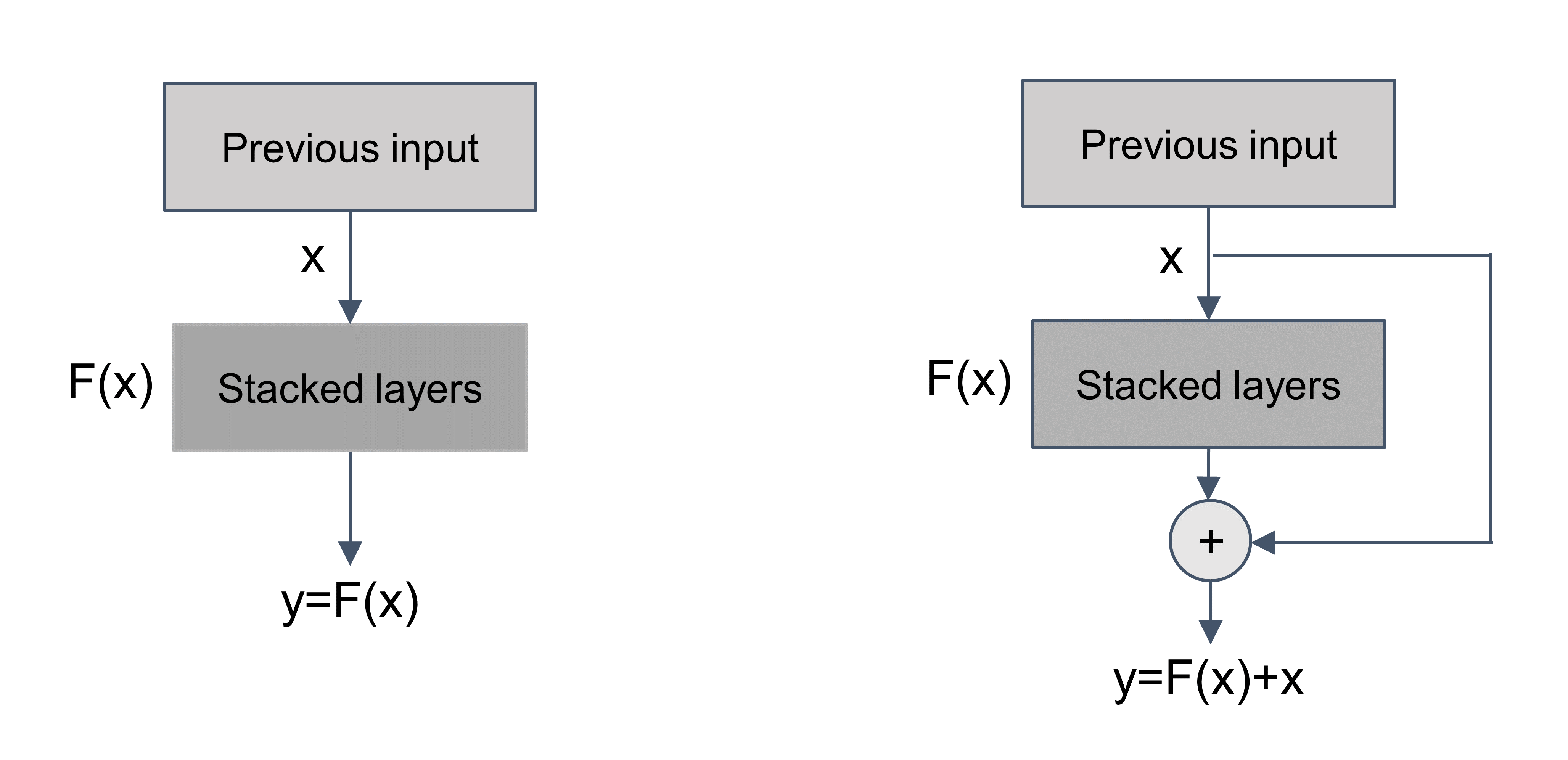}
\caption{ shows an illustration of a typical ResNet’s architecture.}
\label{fig_1}
\end{figure}

Fig. \ref{fig_1} illustrates a typical ResNet architecture. The architecture successfully facilitates message delivery between multiple networks by introducing the residual structure. This enhances the stability of the training process and effectively mitigates the issues of gradient vanishing and Gradient explosion.

\section{Methodology}
In practical applications, models for identifying power quality disturbance signals face various challenges. Signals acquired from real-world scenarios often encounter substantial noise interference, which substantially impairs the capacity of neural networks to learn features. Convolutional kernels and pooling kernels act as common local feature extractors in neural networks. However, signal noise can significantly impact the networks' ability to learn features. Consequently, the performance of numerous existing methods for identifying power quality disturbance signals diminishes notably in such scenarios.

\begin{figure}[ht]
\centering
\includegraphics[width=3.5in]{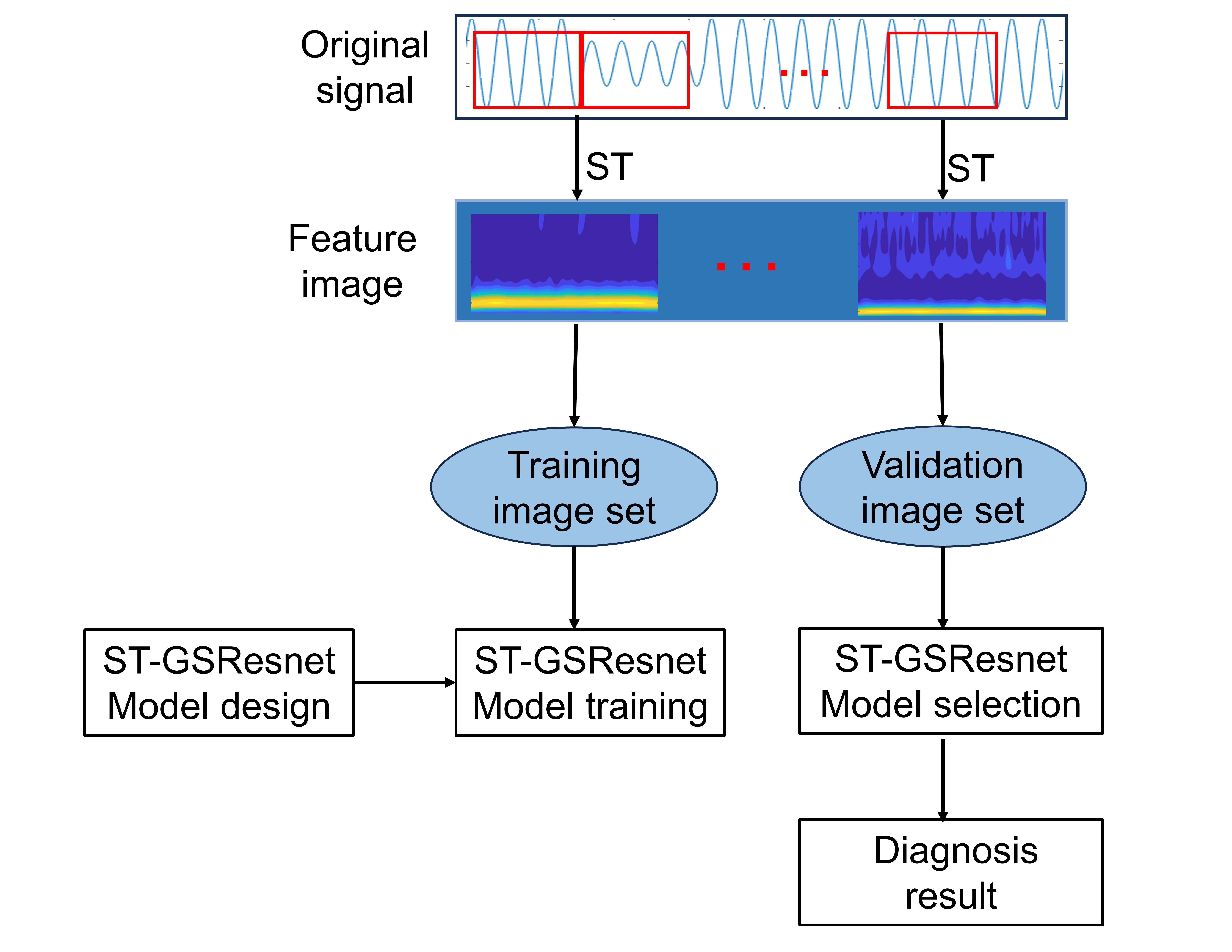}
\caption{ The overall framework of ST-GSResNet.}
\label{fig_2}
\end{figure}
To tackle the aforementioned challenges, this study presents a novel scheme for recognizing PQD signals, termed ST-GSResNet. As shown in Fig. \ref{fig_2}, the proposed method consists of two key components: S-Transform for generating images and GSResNet for classifying images. The specific steps are outlined as follows:

\begin{enumerate}{}{}
\item{The original power quality disturbance signal is transformed using S-Transform to generate its corresponding image representation.}. 
\item{To classify the generated images, the GSResNet is employed as the classifier.}
\item{The model is trained and optimized to enhance recognition accuracy and overall performance.}
\item{The practical application involves feeding the power quality disturbance signal into the model and employing ST-GSResNet for recognition.}
\item{Based on the identification results, suitable measures are implemented to ensure the stability and reliability of power quality.}
\end{enumerate}

\subsection{Generation Two-dimensional time-frequency spectrogram}
The generation of data signals constitutes a foundational step in this research, as a substantial amount of data is crucial for obtaining robust training results in deep learning-based models. The MATLAB tool was employed in this study for data signal generation. This section is in adherence to IEEE standards \cite{ieee11592019}, The frequency for all studied PQDs models studied is fixed at 50 Hz, and parameters are randomly generated within specified ranges. For a comprehensive overview of signal models and standard parameters pertaining to single PQD voltages, please consult Table \ref{tab1}.

\begin{table*}[ht]
\centering
\caption{SIGNAL MODEL OF SINGLE POWER QUALITY DISTURBANCE}
\label{tab1}
\begin{tabular}{l l p{7.9cm} p{4.8cm}}
\toprule
Labels & PQD types & Numerical Model & Parameters \\
\midrule
$\begin{aligned} \text{V1} \\ & \\ & \end{aligned}$ & $\begin{aligned} \text{Harmonic} \\ & \\ & \end{aligned}$ & $\begin{aligned} &V(t)=\sin (\omega t)+\alpha_3 \sin (3 \omega t+\varphi_3)+\alpha_5 \sin (5 \omega t+\varphi_5)+\\&\alpha_7 \sin (7 \omega t+\varphi_7) \\& \end{aligned}$ & $\begin{aligned}&\alpha_3=0 \sim 0.15, \alpha_5=0 \sim 0.15,\\& \alpha_7=0 \sim 0.15, \varphi_3=0 \sim 2 \pi, \\& \varphi_5=0 \sim 2 \pi, \varphi_7=0 \sim 2 \pi \end{aligned}$  \\
V2   &  sag       &  $\begin{aligned}  V(t)=\left(1-\alpha\left(u\left(t-t_1\right)-u\left(t-t_2\right)\right)\right) \sin (\omega t)      \end{aligned}$    &  $\begin{aligned}\alpha=0.1 \sim 0.9, \quad t_2-t_1=4 T \sim 9 T     \end{aligned}$   \\
V3   &  swell        &   $\begin{aligned}V(t)=\left(1-\alpha\left(u\left(t-t_1\right)-u\left(t-t_2\right)\right)\right) \sin (\omega t)      \end{aligned}$      & $\begin{aligned} \alpha=0.1 \sim 0.9, \quad t_2-t_1=4 T \sim 9 T     \end{aligned}$     \\
V4   &  interrupt        & $\begin{aligned} V(t)=\left(1-\alpha\left(u\left(t-t_1\right)-u\left(t-t_2\right)\right)\right) \sin (\omega t)      \end{aligned}$      &  $\begin{aligned} \alpha=0.9 \sim 0.1, \quad t_2-t_1=4 T \sim 9 T   \end{aligned}$      \\ 
V5   &  flicker        & $\begin{aligned} V(t)=\left(1+\alpha_{\mathrm{f}} \sin (\beta \omega t)\right) \sin (\omega t)      \end{aligned}$      &  $\begin{aligned}\alpha_f=0.3 \sim 0.5, \beta=0.1 \sim 0.4   \end{aligned}$       \\
 $\begin{aligned} \text{V6}\\&\\&\end{aligned}$  &  $\begin{aligned} \text{Oscillatory transient (OT)} \\&\\&\end{aligned}$         &  $\begin{aligned} & V(t)=\sin (\omega t)+\alpha_2 \mathrm{e}^{-\frac{\left(t-t_3\right)}{\tau}} \sin \left\{\omega_n\left(t-t_3\right)\right\} \\&\cdot\left\{u\left(t-t_3\right)-u\left(t-t_4\right)\right\} \\&\end{aligned}$      &  
$\begin{aligned} &\alpha_2=0.1 \sim 0.8, \quad \tau=0.008 \sim 0.04,\\& t_4-t_3=0.05 T \sim 3 T, \\& f_n=300 \sim 900 \mathrm{~Hz}   \end{aligned}$       \\
$\begin{aligned} \text{V7}\\&\end{aligned}$  &  $\begin{aligned} \text{Impulsive transient (IT) }\\&\end{aligned}$       & $\begin{aligned} &V(t)=\sin (\omega t)+\alpha_2 \mathrm{e}^{-\frac{\left(t-t_3\right)}{\tau}}\left\{u\left(t-t_3\right)-u\left(t-t_4\right)\right\}\\&\end{aligned}$          &   $\begin{aligned}&\alpha_2=1 \sim 10, \tau=0.008 \sim 0.04, \\&t_4-t_3=0.05 T \sim 3 T  \end{aligned}$  \\
\bottomrule
\end{tabular}
\end{table*}

\begin{figure}[ht]
\centering
\includegraphics[width=3.5in]{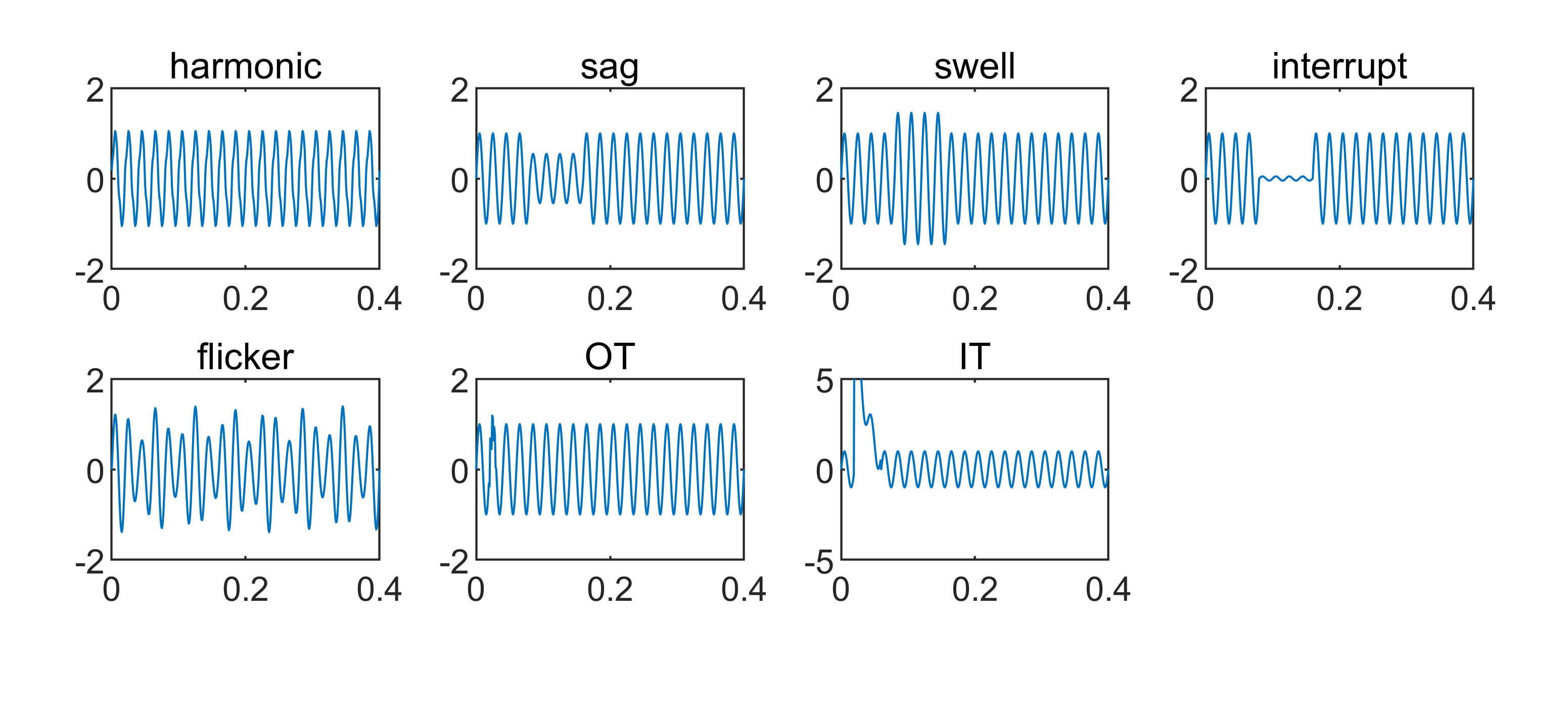}
\caption{ Typical waves of 7 single PQDs. }
\label{fig_3}
\end{figure}

Utilizing the previously mentioned signal model, a MATLAB simulation script enables the generation of a substantial quantity of individual disturbance samples.Fig. \ref{fig_3} illustrates the typical waveforms.

This section selects prevalent PQDs along with their corresponding signal models and parameter explanations. These encompass 7 types of dual-disturbance power quality compound disturbances. The fundamental frequency for all the investigated models of PQ disturbances was set at 50 Hz, with parameters being randomly generated within the specified ranges. Table \ref{tab:tab2} delineates the signal models and standard parameters for the mixed PQDs.

\begin{table}[ht]
\caption{ENVIRONMENT REQUIRED FOR THE EXPERIMENT}
\label{tab:tab2}
\centering
\begin{tabular}{l l}
\toprule[1pt] 
Labels  &  PQD types  \\
\midrule
V8 & harmonics+sag  \\
V9 & harmonics+swell \\
V10 & interruption + harmonics \\
V11 & Impulsive transient + sag  \\
V12 & Impulsive transient + swell \\
V13 & Impulsive transient + flicker  \\
V14 & Impulsive transient + harmonics \\
V15 & harmonics + Oscillatory transient + sag  \\
V16 & harmonics + Oscillatory transient + swell \\
V17 & flicker+ Impulsive transient  + harmonics \\
V18 & harmonics + Oscillatory transient + Impulsive transient +sag \\
\bottomrule[1pt] 
\end{tabular}
\end{table}

\begin{figure}[ht]
\centering
\includegraphics[width=3.5in]{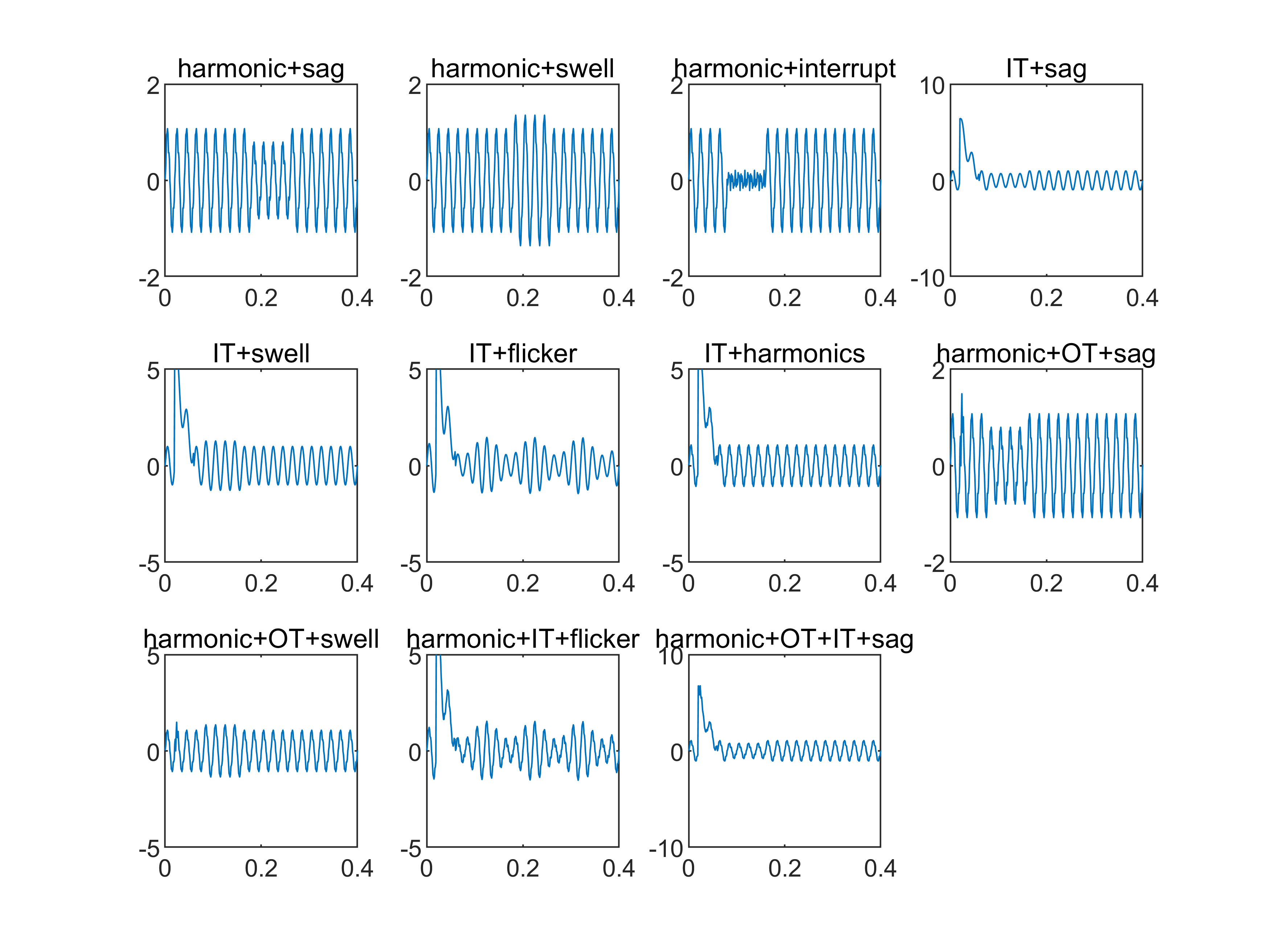}
\caption{ Typical waves of 11 mixed PQDs.}
\label{fig_4}
\end{figure}

Utilizing the previously mentioned mixed disturbance signal models, a MATLAB script can generate a substantial number of mixed disturbance samples. Figure 5 illustrates the representative waveforms. The typical waveform is then transformed into a time-frequency plot, as shown in Fig. \ref{fig_4}.

After generating, validating, and collecting the data, the subsequent step is to create a scalogram that characterizes the voltage disturbances in the collected signals. Within the MATLAB environment, time-frequency conversion of the generated signal waveforms is performed using the S-Transform to produce their corresponding spectrograms.

\begin{figure}[ht]
\centering
\includegraphics[width=3.5in]{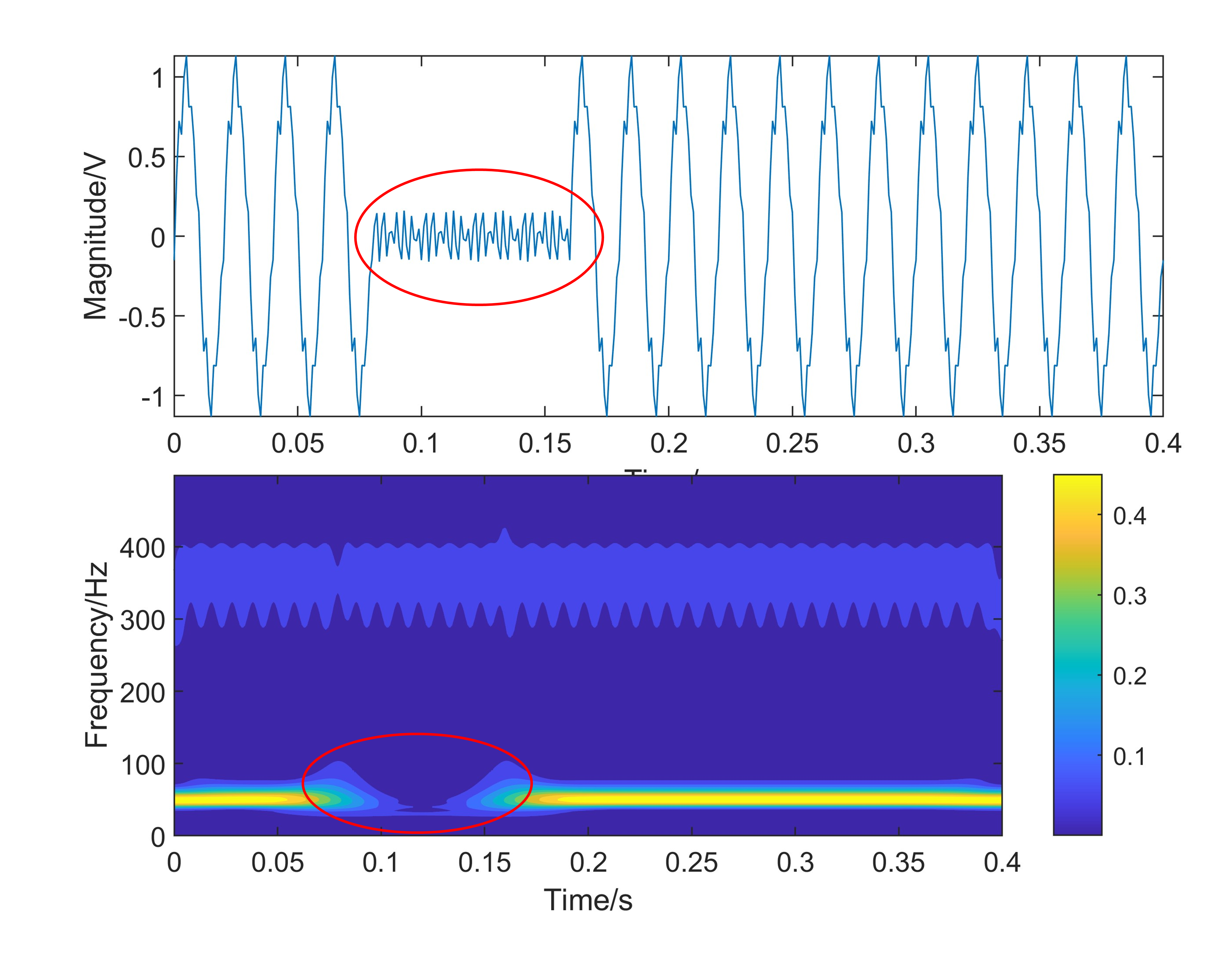}
\caption{ The time-frequency plot of the signal V10, generated by the S-Transform, is compared to the original waveform.}
\label{fig_5}
\end{figure}

By utilizing the coefficients generated by the S-Transform, MATLAB's color mapping functions, such as the "jet 264" style, can be employed to analyze these signals and observe the energy scales in the scalogram. This process involves applying the absolute values of the coefficients. Fig. \ref{fig_5} illustrates the time-frequency plot of the signal V10 generated by the S-Transform compared to the original waveform.

After generating all the images containing only scalar plots, they are saved in specific folders corresponding to each PQ disturbance category. The saved images are formatted to be 240x240 pixels and saved in the PNG format using the "imwrite" command and corresponding supporting code. To achieve this, the "imwrite" command and corresponding supporting code are used to save a large number of images. The generated dataset consists of 18 categories, each containing 1000 images. Subsequently, these images are divided into training and testing sets in a 7:3 ratio for model evaluation.

\subsection{Improved ResNet}
\begin{figure}[ht]
\centering
\includegraphics[width=3.5in]{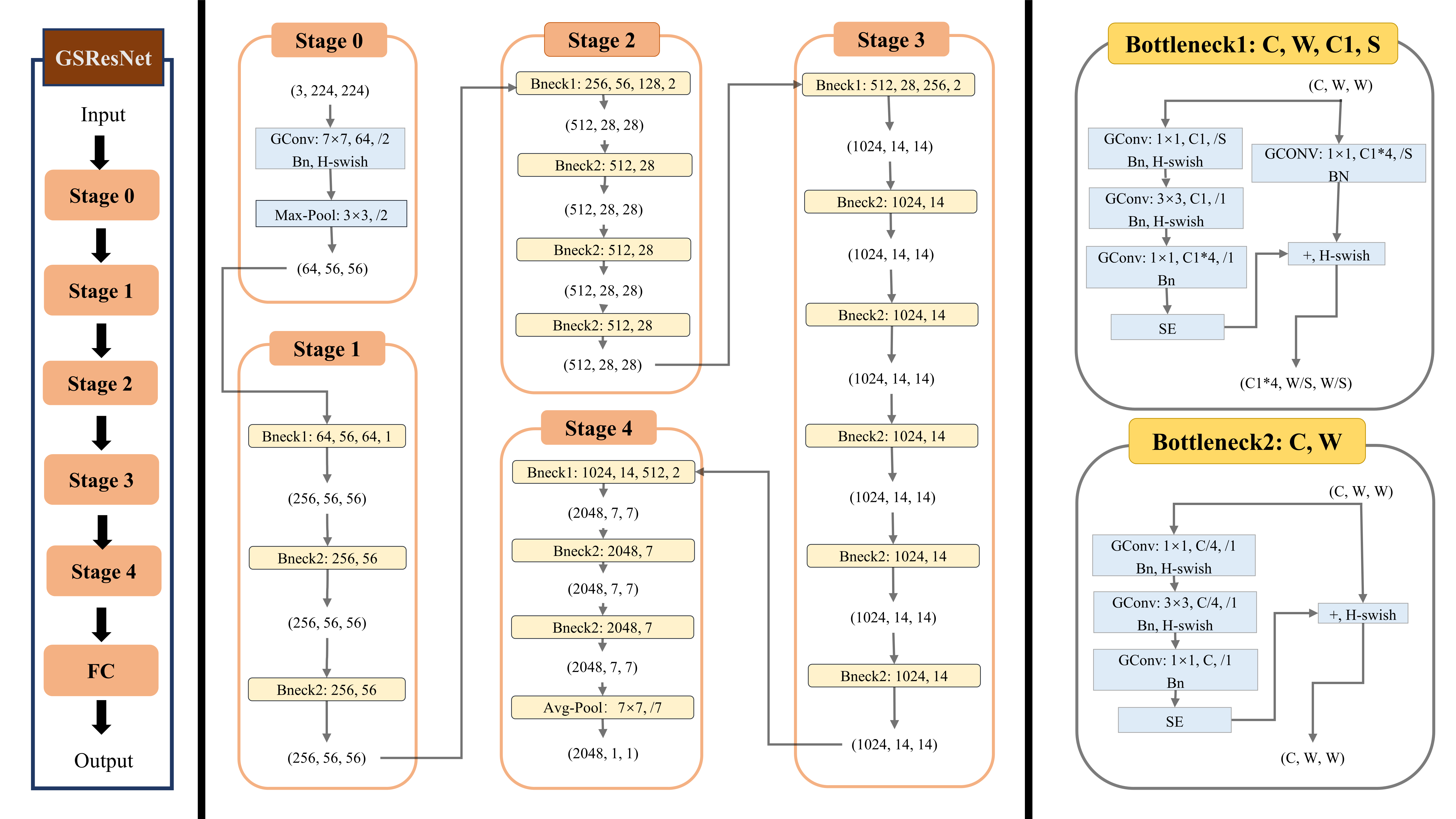}
\caption{ illustrates the structure of the proposed GSResNet model.}
\label{fig_6}
\end{figure}

The improved ResNet model will be used to classify the time-frequency spectrograms of the 18 classes of power quality disturbance signals, which were previously generated by utilizing the S-Transform.As revealed in Fig. \ref{fig_6}, the model diagram presents the overall architecture of our proposed model, encompassing the general design on the left, the specific structure of each stage in the middle, and the bottleneck structure on the right. In this context, C, H, W, and S denote the number of channels, height, width, and stride, respectively.

\textit{1) Group convolution: }The significance of convolutional neural networks in image processing is undeniable. Convolutional operations boast robust feature extraction capabilities and require fewer parameters compared to fully connected networks. Convolution inherently thrives when processing two-dimensional structured data, particularly images. The concept of grouped convolution was first presented in AlexNet\cite{krizhevsky2012} in 2012.Duringthattime, considering the memory and computational constraints of a single GPU, grouped convolution was implemented by dividing the feature map and convolution kernel into 'g' groups along the channel direction. The results obtained from each convolution group are then concatenated to yield the final outcome. 

Relative to the traditional convolution operation employed in the ResNet model, grouped convolution augments the network's ability to model non-linearity and extract intricate features. Additionally, due to the parallelisability of convolution operations, grouped convolution bolsters the computational efficiency of the network. Consequently, this study substitutes the original convolution operation with grouped convolution, thereby further amplifying the capabilities of the original ResNet model.

\begin{figure}[t]
\centering
\subfloat[]{\includegraphics[width=0.7\linewidth]{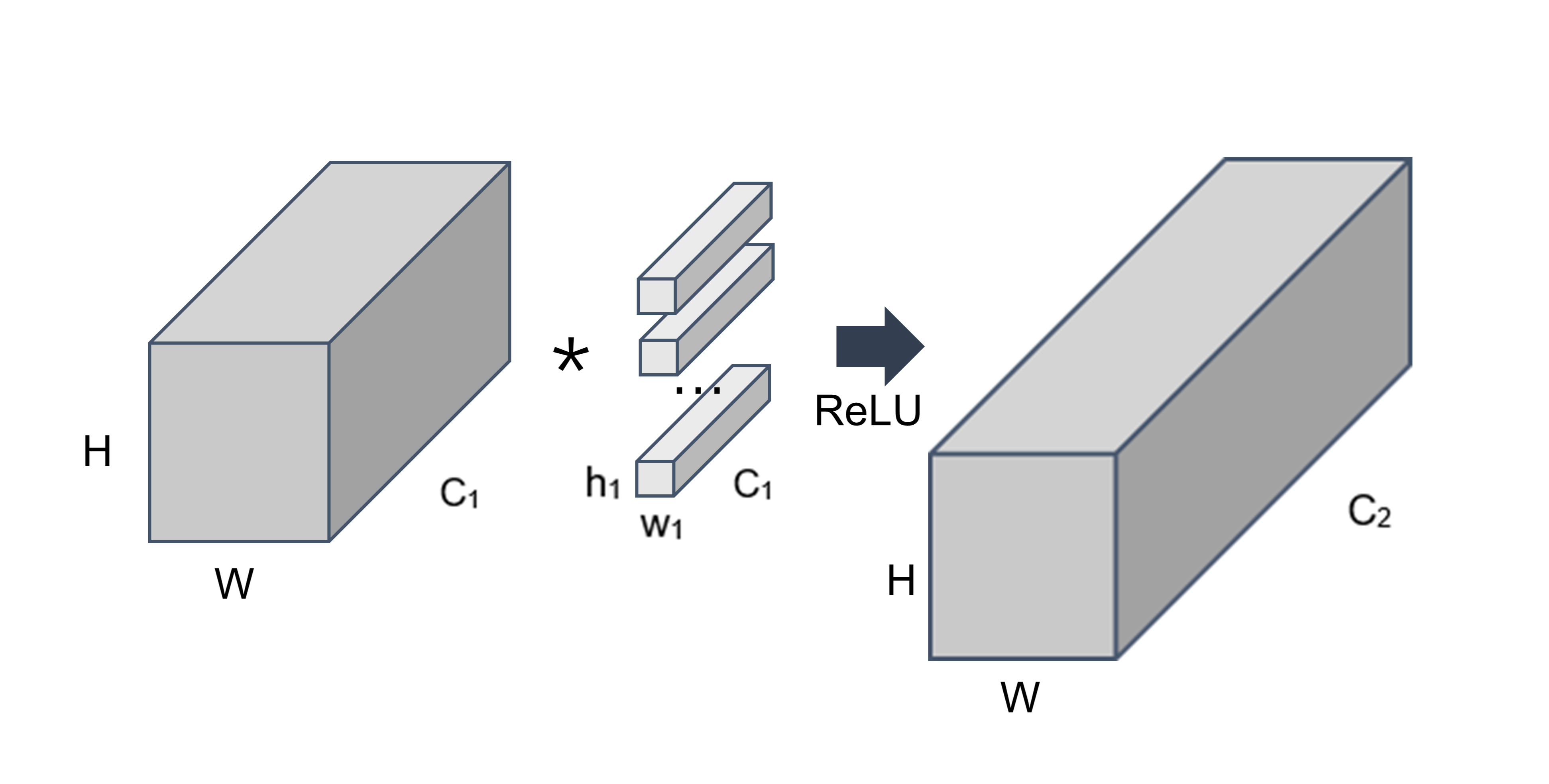}%
\label{fig_first1_case}}

\subfloat[]{\includegraphics[width=0.7\linewidth]{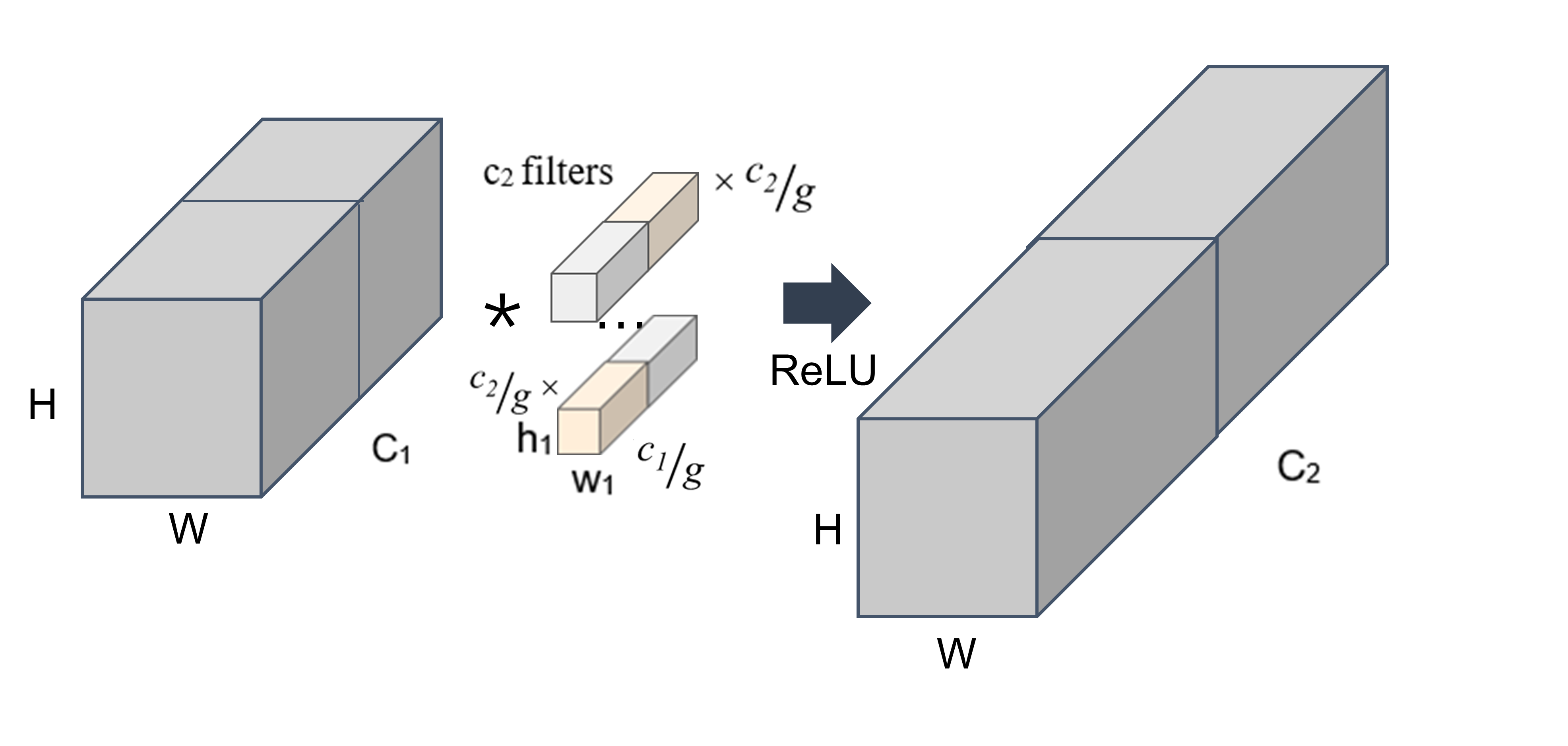}%
\label{fig_second1_case}}
\caption{ The figure above illustrates a grouped convolution with two sets of filters. In each of the sets of filters depicted in panel b, every filter has a depth that is half that of the nominal 2D convolution shown in panel a. (a) Convolution. (b) Convolution with filter groups.}
\label{fig_7}
\end{figure}

The standard convolution operation is a fundamental neural network operation used to extract feature information from an input feature map. Illustrated in the figure, this operation involves element-wise multiplication of a set of learnable convolution kernels element-wise with the input feature map and summing the results to generate the output feature map. Grouped convolution involves dividing the input feature map into "g" groups along the channel dimension and partitioning the convolution kernel into "g" groups accordingly. Consequently, the size of the convolution kernel in each group is adjusted, resulting in the output feature map being divided into "g" groups along the channel dimension as well. Fig. \ref{fig_7} show the comparison between ordinary convolution and grouped convolution.

\textit{2) Squeeze-and-Excitation: }The core of the SE module \cite{hu2018} is the squeeze-and-excitation block. The SE module employs a Multi-Layer Perceptron (MLP) to model each channel, thereby concurrently producing a weight vector that signifies channel attention. Specifically, the feature maps are converted into an intermediate vector through a fully connected layer, which is then followed by the introduction of nonlinearity via an element-wise ReLU activation function. Lastly, the intermediate vector is transformed into a scalar value through another fully connected layer, and the output is interpreted as a vector of channel weights through the application of a sigmoid function.

\begin{figure}[ht]
\centering
\includegraphics[height=1.5in,width=3in]{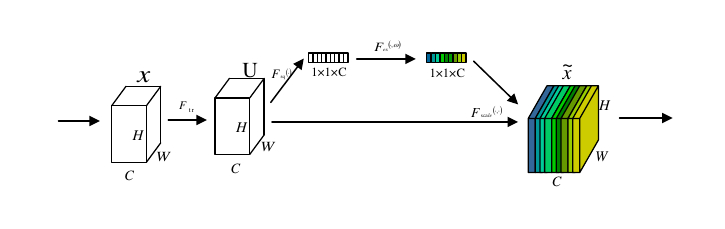}
\caption{ The architecture of the SE module add to ResNet.}
\label{fig_8}
\end{figure}

The complete SE module is composed of the combination of Squeeze and Excitation components. The insertion of the SE block into the network's convolutional layers aids in the representation of feature correlations and differentiations, thereby leading to the automatic adjustment of channel significance. The integration of the SE block resulted in a significant enhancement in the model's image classification performance. The SE module was designed based on a straightforward yet effective concept, allowing it to be integrated into various CNN architectures. Fig. \ref{fig_8} illustrates the architecture of the SE module when integrated into ResNet.

Fundamentally, the SE module plays a pivotal role in enhancing ResNet's performance. The integration of the SE module allows networks to flexibly select and weight features, subsequently enhancing the model's capacity for representation and classification. In the context of image classification, the SE module effectively steers the network towards the most pertinent features, thereby facilitating learning and accurately discerning complex cases. Moreover, due to its highly flexible and scalable attributes, the SE module can be seamlessly integrated into existing ResNet network architectures. The SE module can enhance network performance by incorporating a limited number of computational operations without introducing additional parameters. The incorporation of the SE module enables the achievement of improved model performance with a minimal increase in computational burden.

\textit{3) Nonlinearities: }In the seminal work of \cite{koonce2021}, an innovative nonlinearity termed as h-swish was introduced, showcasing significant effectiveness when flawlessly integrated as a substitute for ReLU. This integration significantly enhances the precision of neural networks. The nonlinearity is explicitly defined as:

\begin{equation}
\label{deqn_ex1}
\text{h-swish}(x) = x \times \text{ReLU}\left(\frac{x+3}{6}\right)
\end{equation}

In previous studies, a nonlinearity termed swish was presented, which, when utilized as a direct replacement for ReLU, offers numerous advantages. The nonlinearity is defined as:

\begin{equation}
\label{deqn_ex1}
\text{Swish}(x) = x \cdot \sigma(x) 
\end{equation}
\begin{equation}
\label{deqn_ex1}
\sigma(x) = \frac{1}{1 + e^{-x}}
\end{equation}

The benefits of using the sigmoid function in neural network layers include its unbounded nature, smoothness, and non-monotonic properties, which together enhance the network's expressive power. However, its nonlinearity, while improving accuracy, significantly increases computational demands, particularly on mobile devices. To balance model accuracy and computational efficiency, we employ the h-swish activation function.In the swish function, the input value, x, is scaled by the sigmoid function and then multiplied by x. This setup allows the function to approximate linearity for large x values and display a strong non-linear behavior for small x values.In contrast, the h-swish function normalizes the input value, x, mapping it to the [0, 1] interval, before multiplying it by the original input, x. This approach results in a piecewise linear approximation, making the function simpler and more computationally efficient for both small and large x values.

\section{Experimental}
\subsection{Experimental Setup}

\textit{1) Experimental Environment and Training Strategy for Model: } During training, we employ the Nadam optimization algorithm with a first-order momentum of 0.9, a second-order momentum of 0.999, and a weight decay parameter of 1e-7. We set the initial learning rate to 0.0001 utilizing a cosine annealing learning rate strategy. This allows the learning rate to fluctuate during training according to a cosine function, while maintaining an overall downward trend. The default batch size is set to 16, and training spans 100 epochs, taking approximately 2 to 3 hours.All experiments were conducted on the dataset we created. The detailed training configuration is shown in Table \ref{tab:tab3}.

\begin{table}[ht]
\caption{ENVIRONMENT REQUIRED FOR THE EXPERIMENT
\label{tab:tab3}}
\centering
\begin{tabular}{l c c c}
\toprule[1pt] 
Laboratory Setting  &  Configuration Information  \\
\midrule
CPU   &  Intel Core i5-13490F            \\
GPU   &  Nvidia GeForce RTX 4060ti GPU 16G             \\
CUDA   &  11.6              \\
Running System  &  Ubuntu 22.04            \\ 
Programming Language & Python 3.8\\
Deep Learning Framework & PyTorch 1.13.1\\
\bottomrule[1pt] 
\end{tabular}
\end{table}

\textit{2) Evaluation metrics: }Evaluation metrics are crucial for assessing the effectiveness of diagnostic algorithms, rendering them essential in data analysis. In the context of intelligent fault diagnosis, the accuracy rate serves as a widely adopted evaluation metric. It measures the proportion of accurate predictions in the total sample and thus indicates the classifier's performance. Higher accuracy rates typically indicate enhanced classifier performance. The definitions of indicators are provided in Eq. (\ref{deqn_ex1}). 

\begin{equation}
\label{deqn_ex1}
\text{Accuracy} = \frac{TP + TN}{TP + TN + FP + FN}
\end{equation}

The terms TP, TN, FP, and FN correspond to true positives, true negatives, false positives, and false negatives, respectively, within the context of class \textit{i}. In the context of our study, represents the output of the network model, while denotes the true label.
Accuracy is an intuitive metric that represents the proportion of samples correctly predicted by a model, typically expressed as a percentage. This characteristic renders accuracy a highly intuitive evaluation metric that effectively communicates the model's overall performance. While accuracy is an intuitive and essential metric, relying solely on it to evaluate the performance of an algorithmic model requires more scientific rigor and comprehensiveness. Therefore, this paper additionally employs a confusion matrix to comprehensively assess the classifier's performance. The confusion matrix offers more information than a single accuracy measure and can depict the confusion between different categories, including the proportions of correctly and incorrectly classified cases. Utilizing the confusion matrix, we can compute the number of correct and incorrect judgments made by the model on the samples. 

\begin{figure}[!t]
\centering
\subfloat[]{\includegraphics[width=0.45\linewidth]{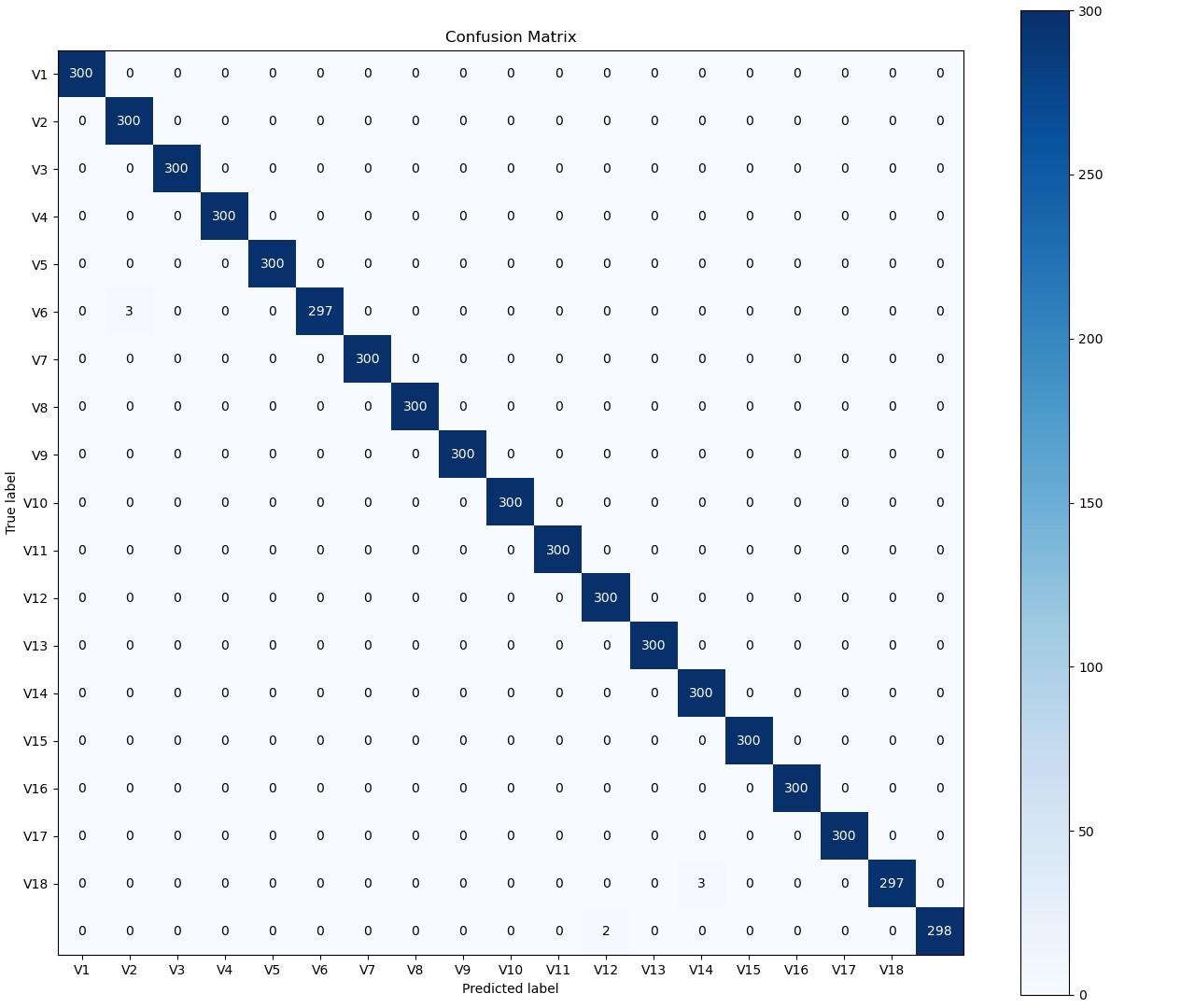}%
\label{fig:noiseless}}
\hfill
\subfloat[]{\includegraphics[width=0.45\linewidth]{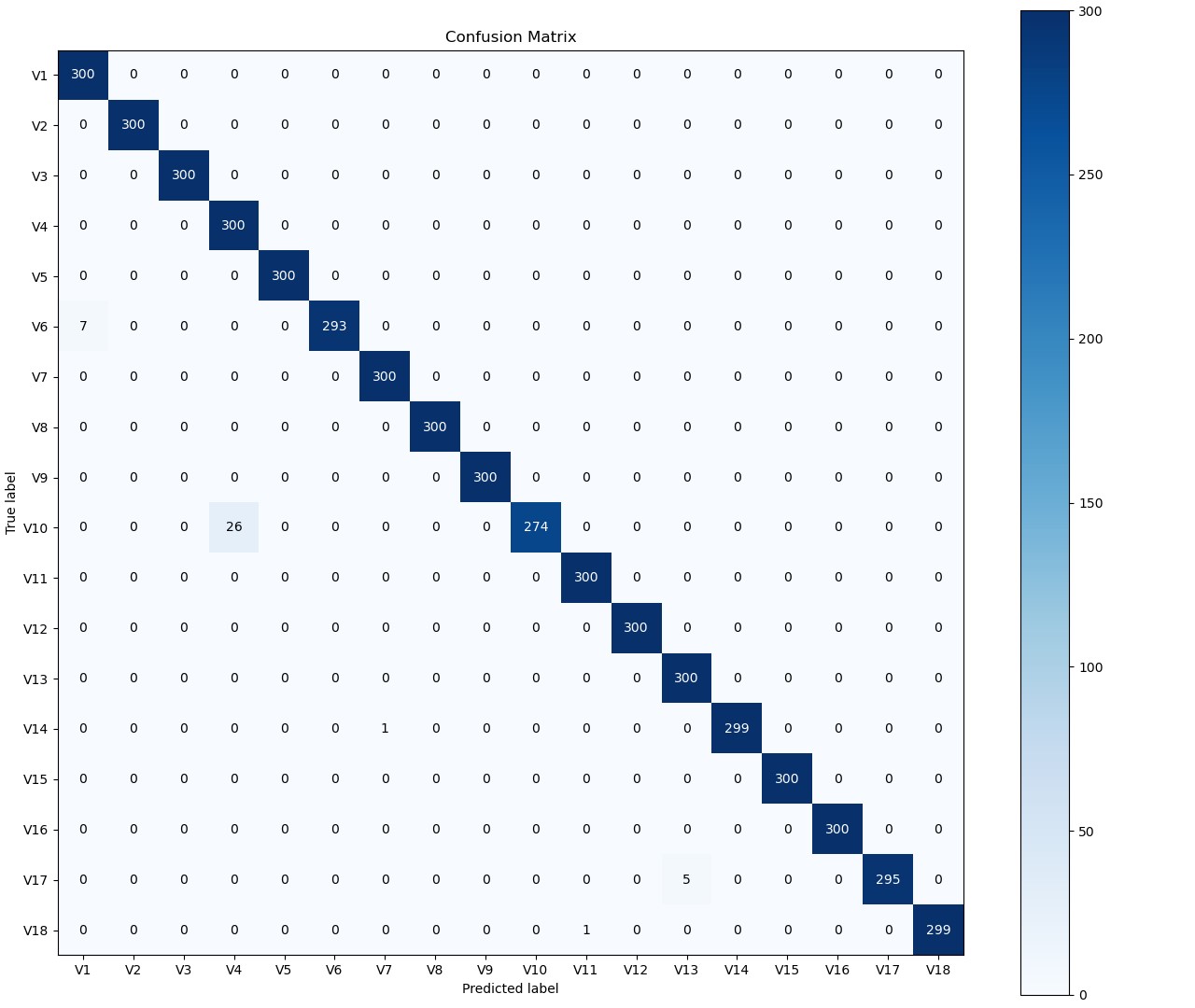}%
\label{fig:snr40}}

\subfloat[]{\includegraphics[width=0.45\linewidth]{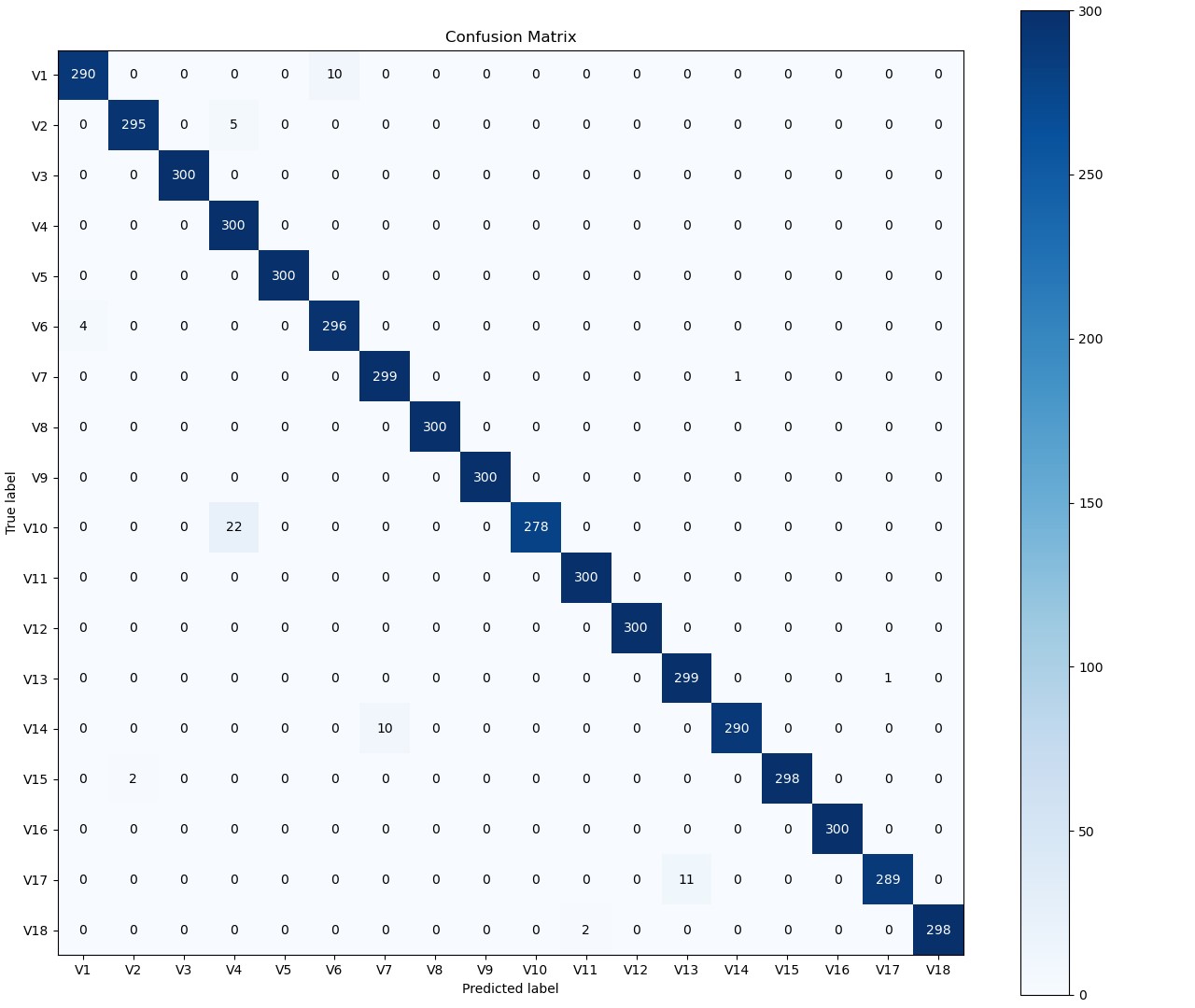}%
\label{fig:snr30}}
\hfill
\subfloat[]{\includegraphics[width=0.45\linewidth]{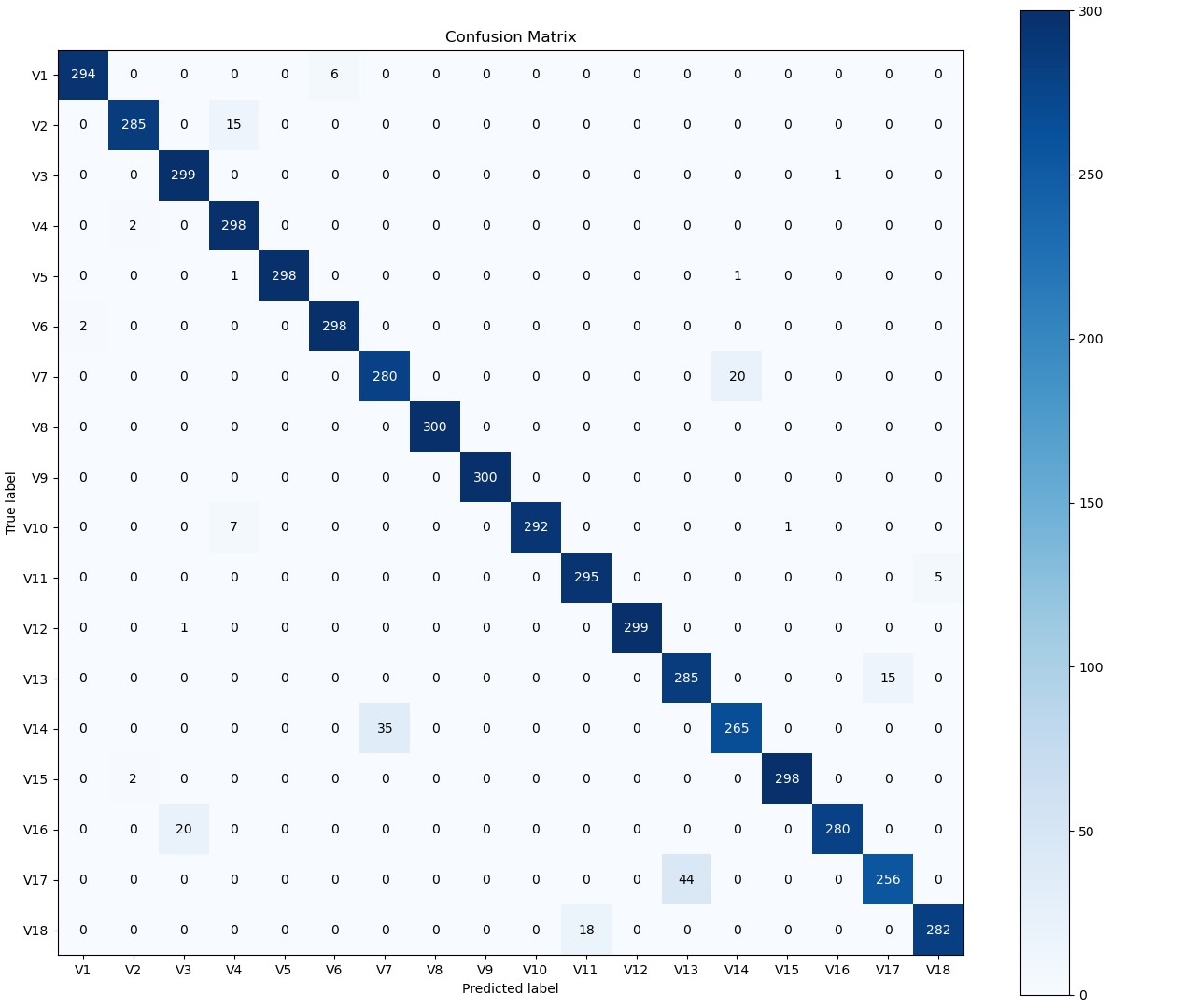}%
\label{fig:snr20}}
\caption{GSResNet Test Results: (a) Noiseless , (b) SNR = 40dB , (c) SNR = 30dB , (d) SNR = 20dB .}
\label{fig:gsresnet}
\end{figure}

\begin{figure*}[!t]
\centering
\subfloat[]{\includegraphics[width=0.24\textwidth]{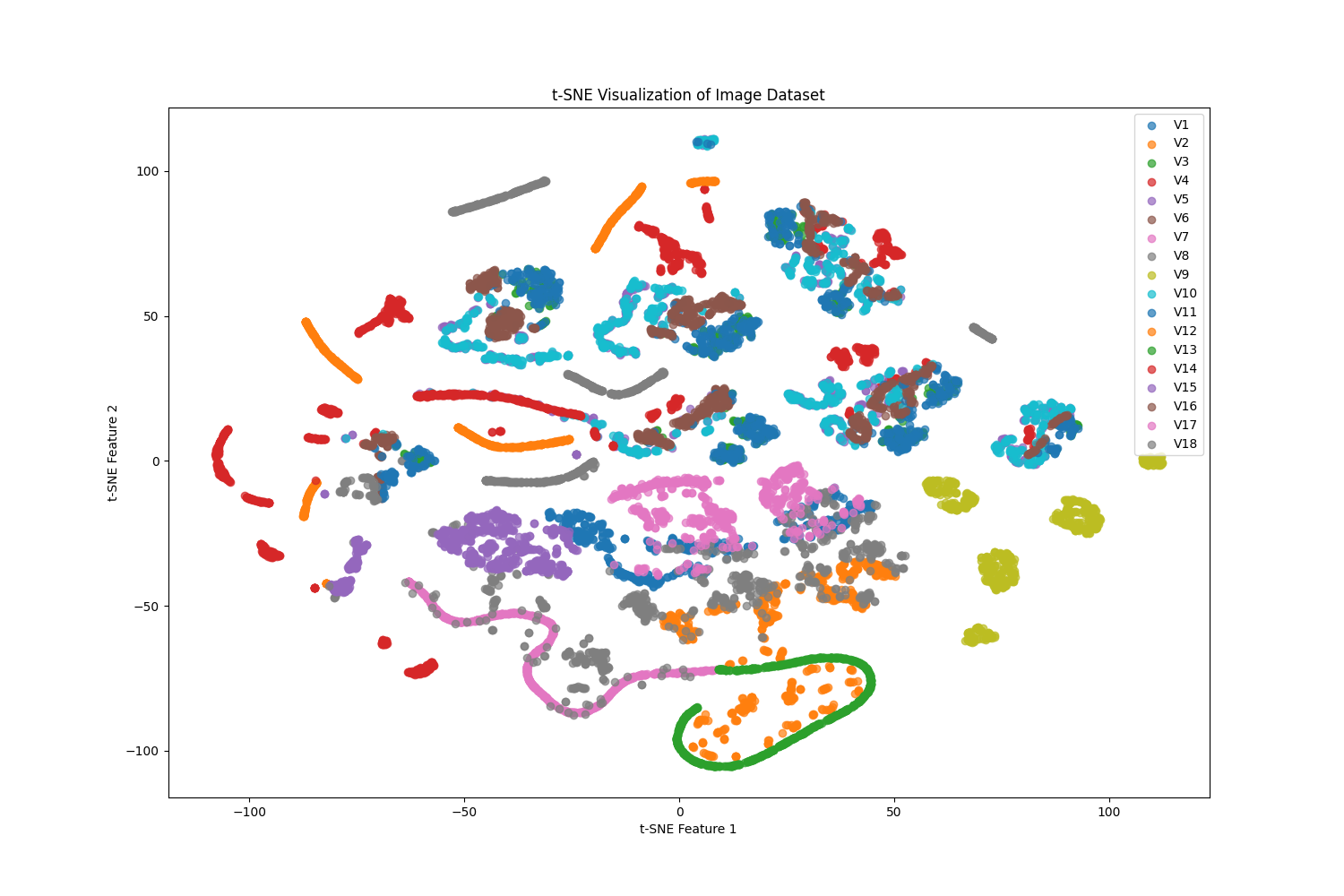}%
\label{fig:noiseless12}}
\hfill
\subfloat[]{\includegraphics[width=0.24\textwidth]{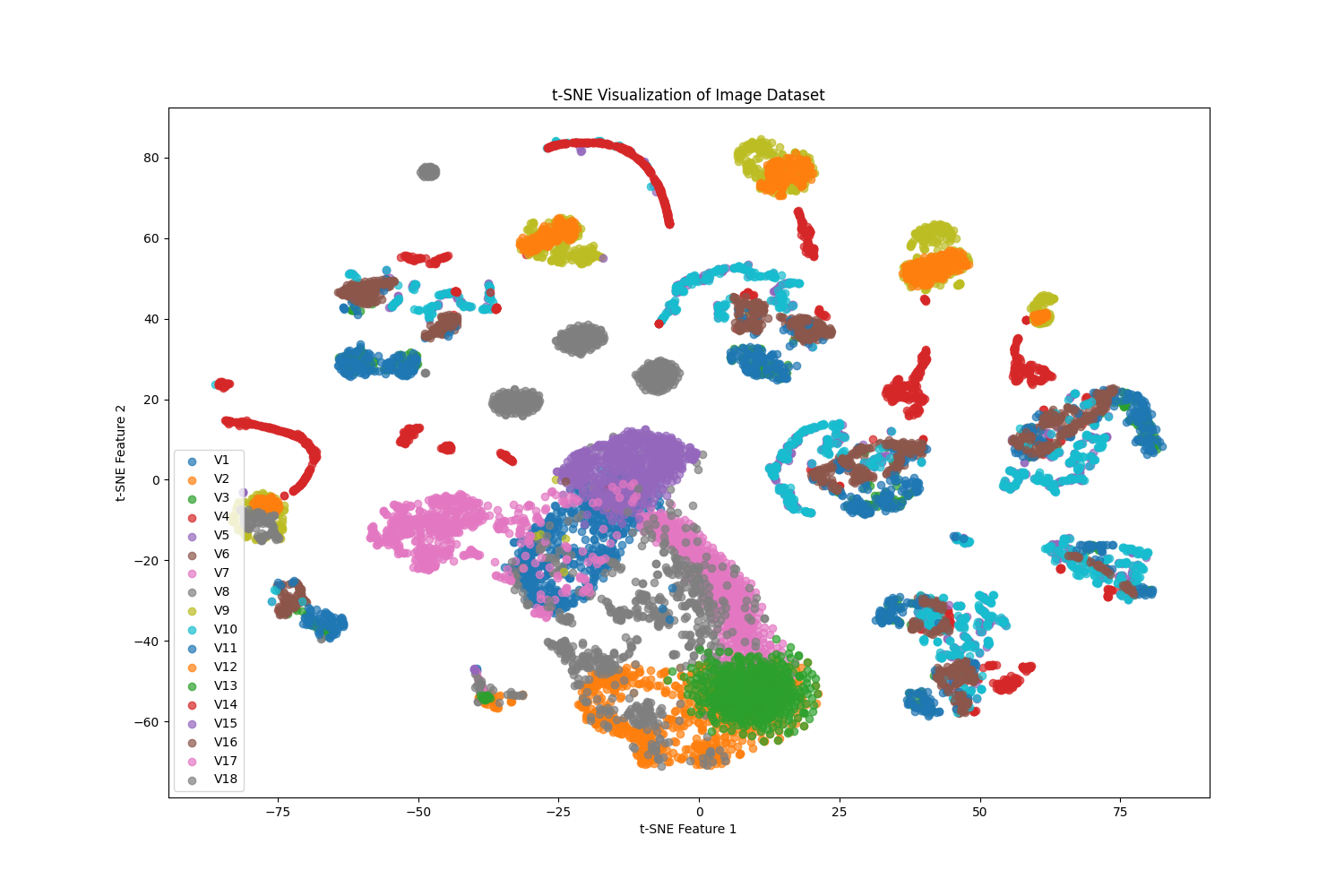}%
\label{fig:snr4012}}
\hfill
\subfloat[]{\includegraphics[width=0.24\textwidth]{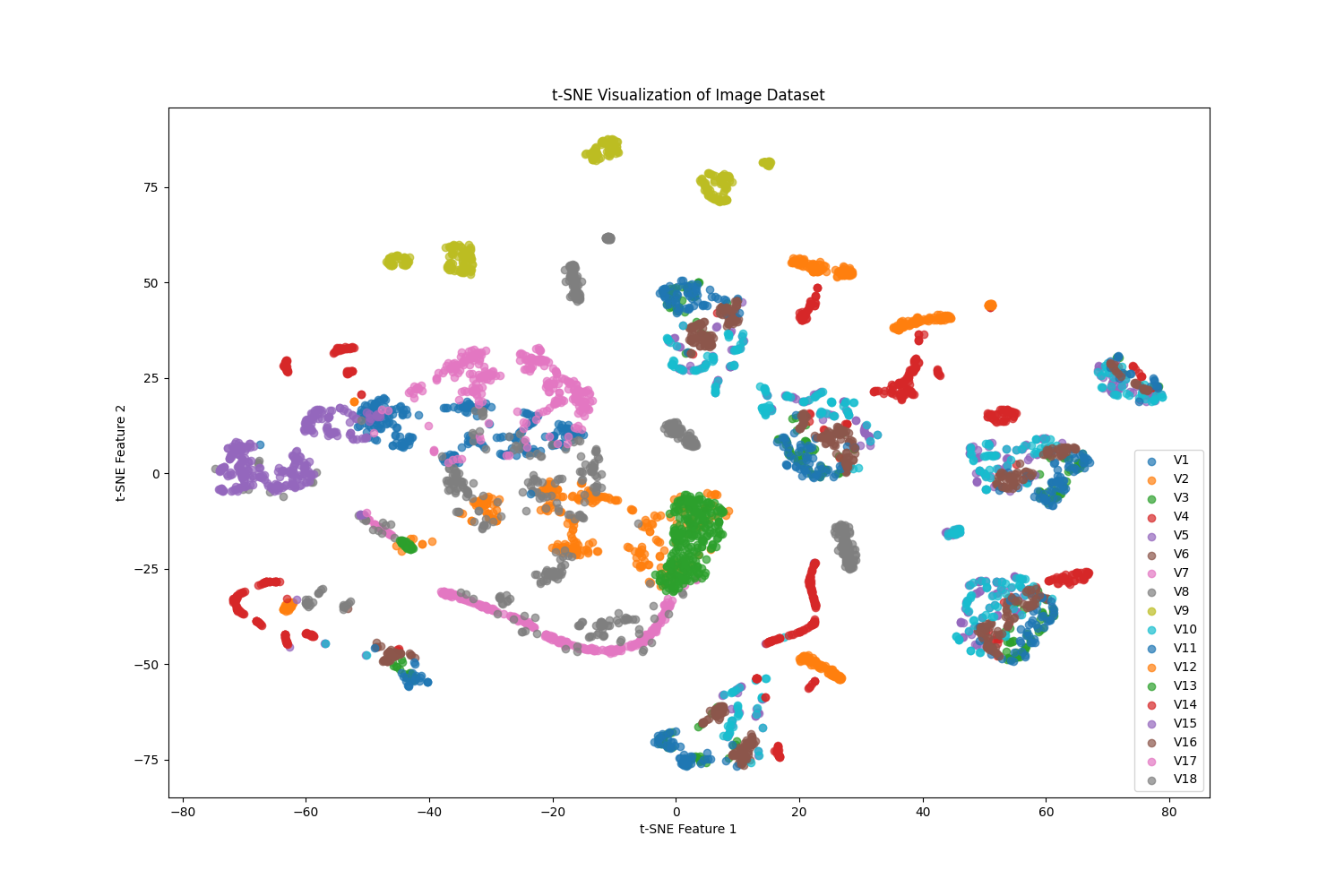}%
\label{fig:snr3012}}
\hfill
\subfloat[]{\includegraphics[width=0.24\textwidth]{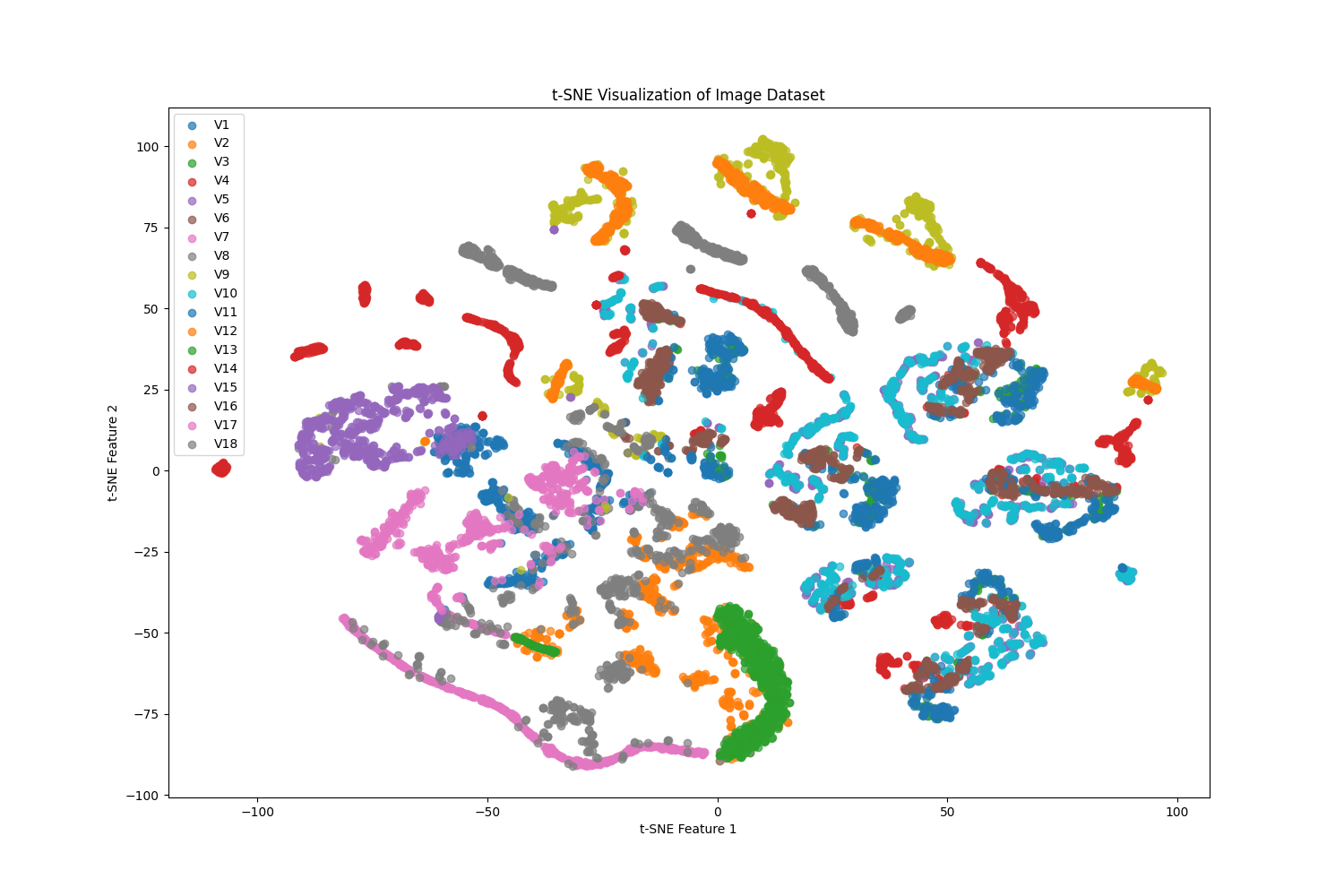}%
\label{fig:snr2012}}

\subfloat[]{\includegraphics[width=0.24\textwidth]{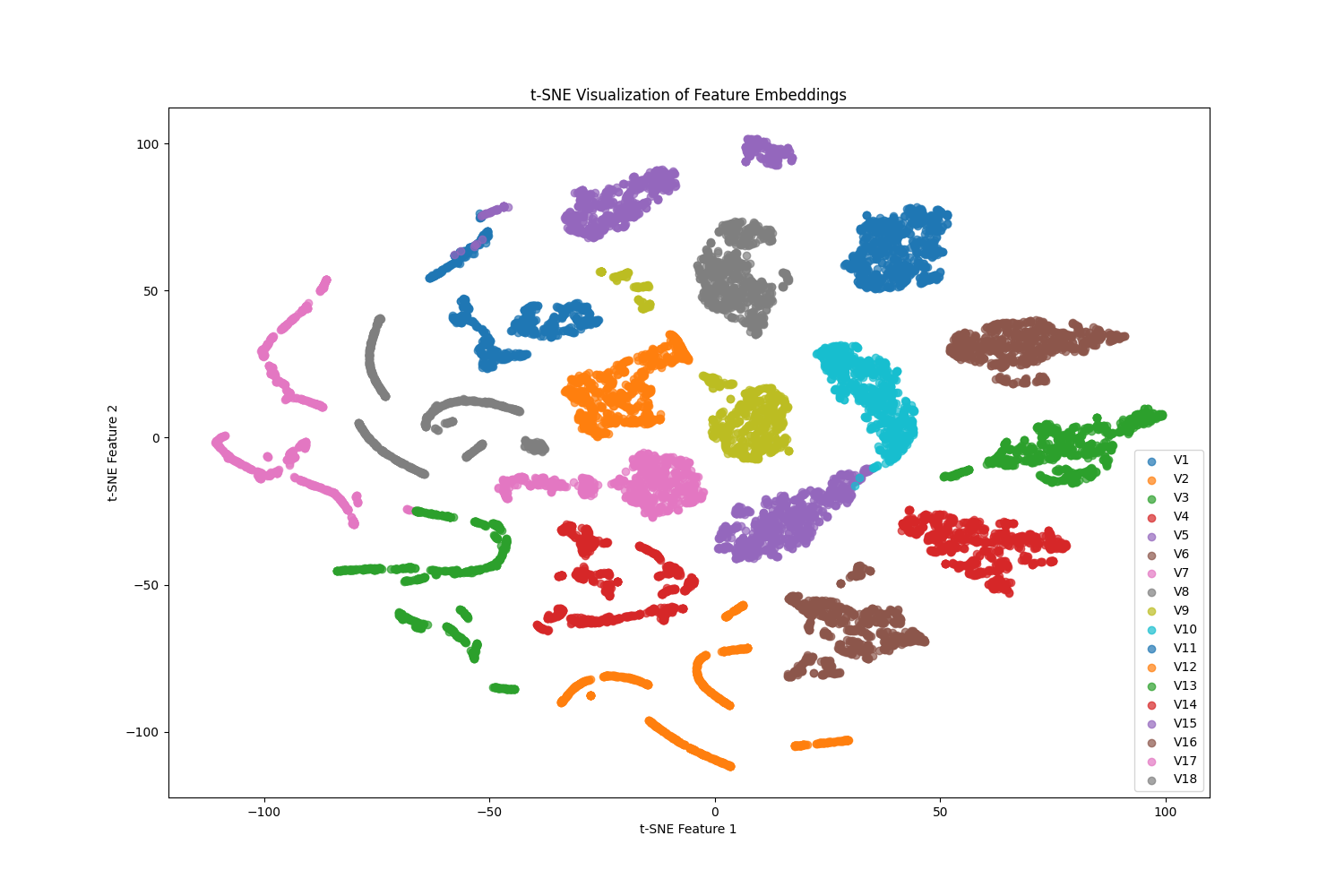}%
\label{fig:noiseless34}}
\hfill
\subfloat[]{\includegraphics[width=0.24\textwidth]{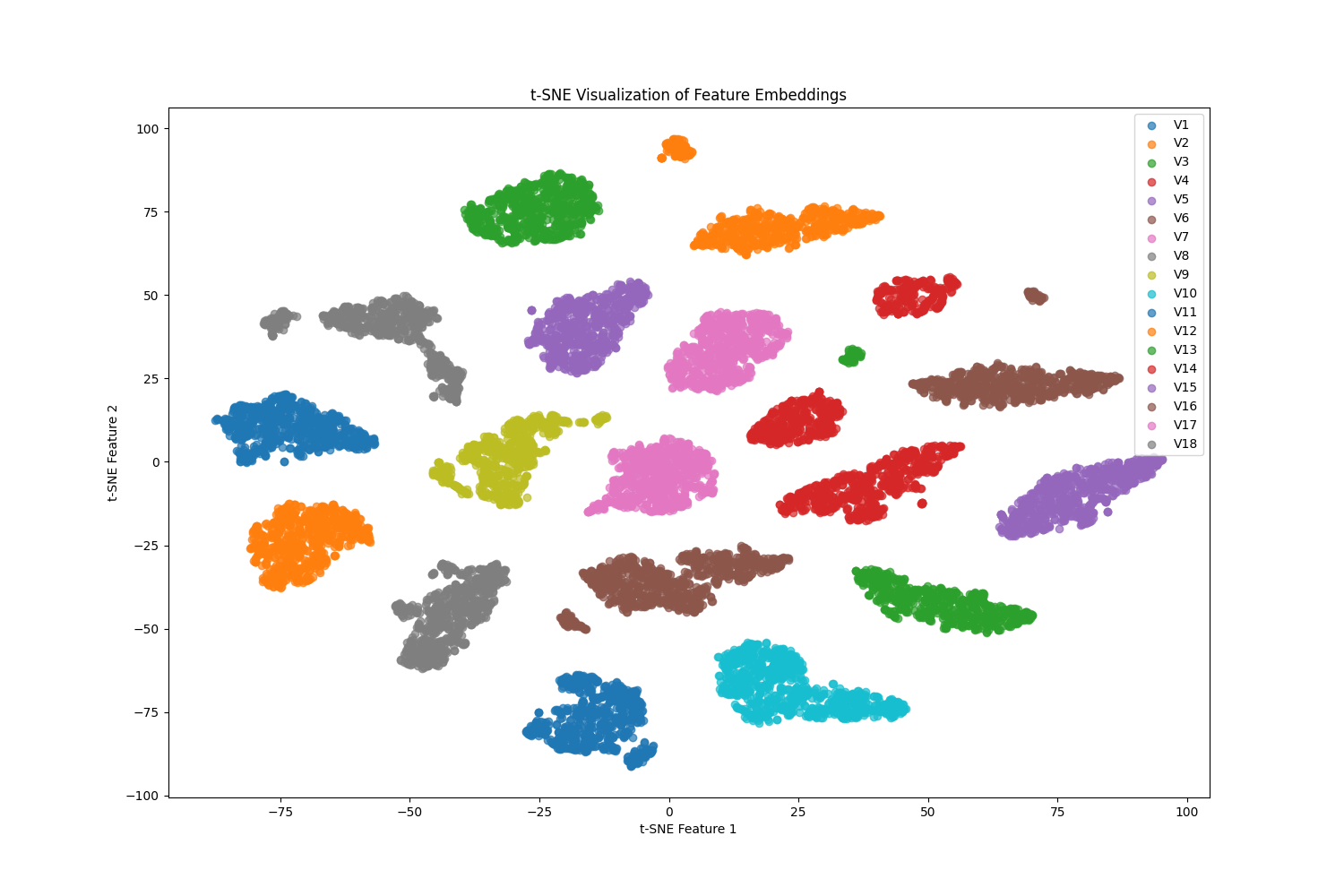}%
\label{fig:snr4034}}
\hfill
\subfloat[]{\includegraphics[width=0.24\textwidth]{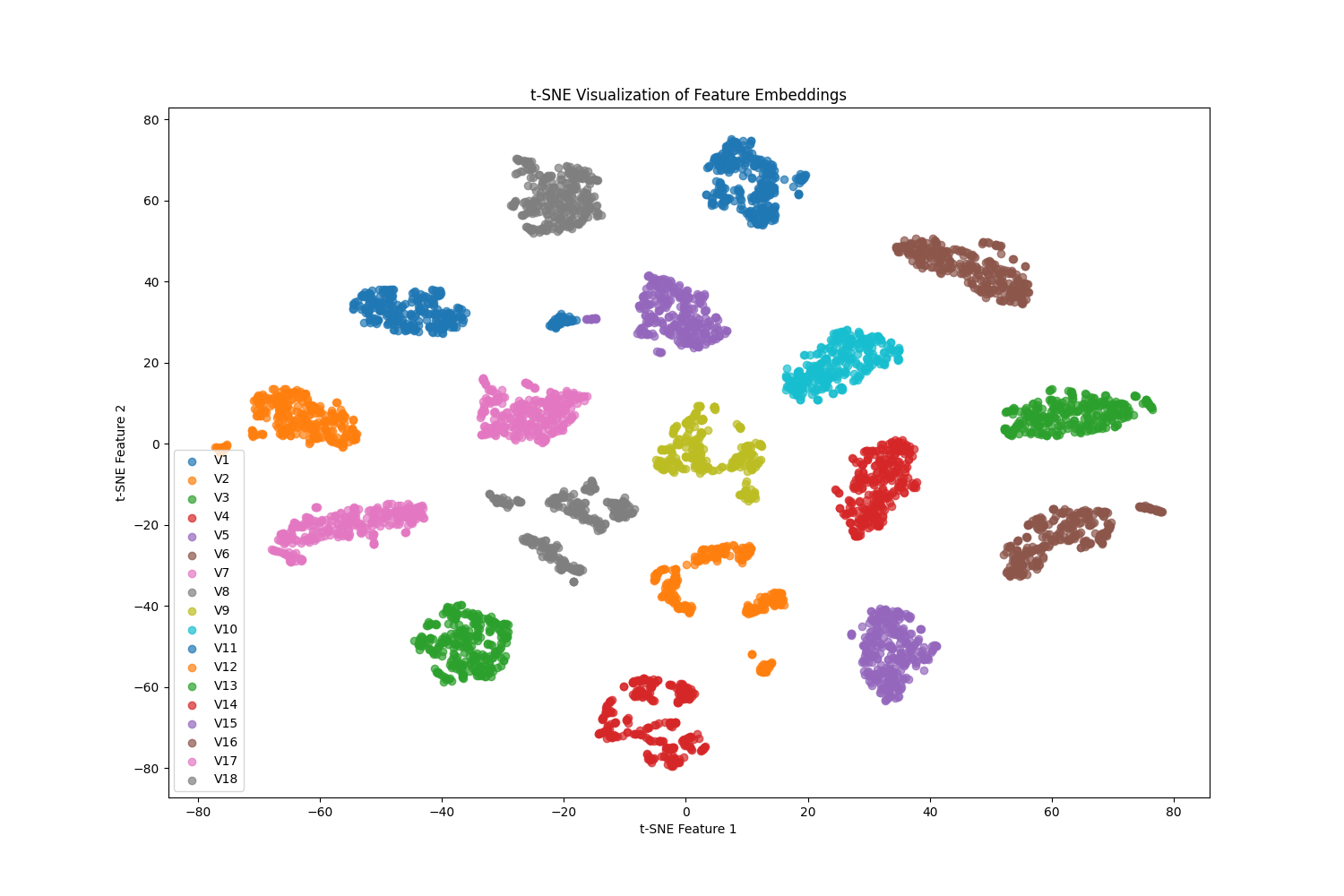}%
\label{fig:snr3034}}
\hfill
\subfloat[]{\includegraphics[width=0.24\textwidth]{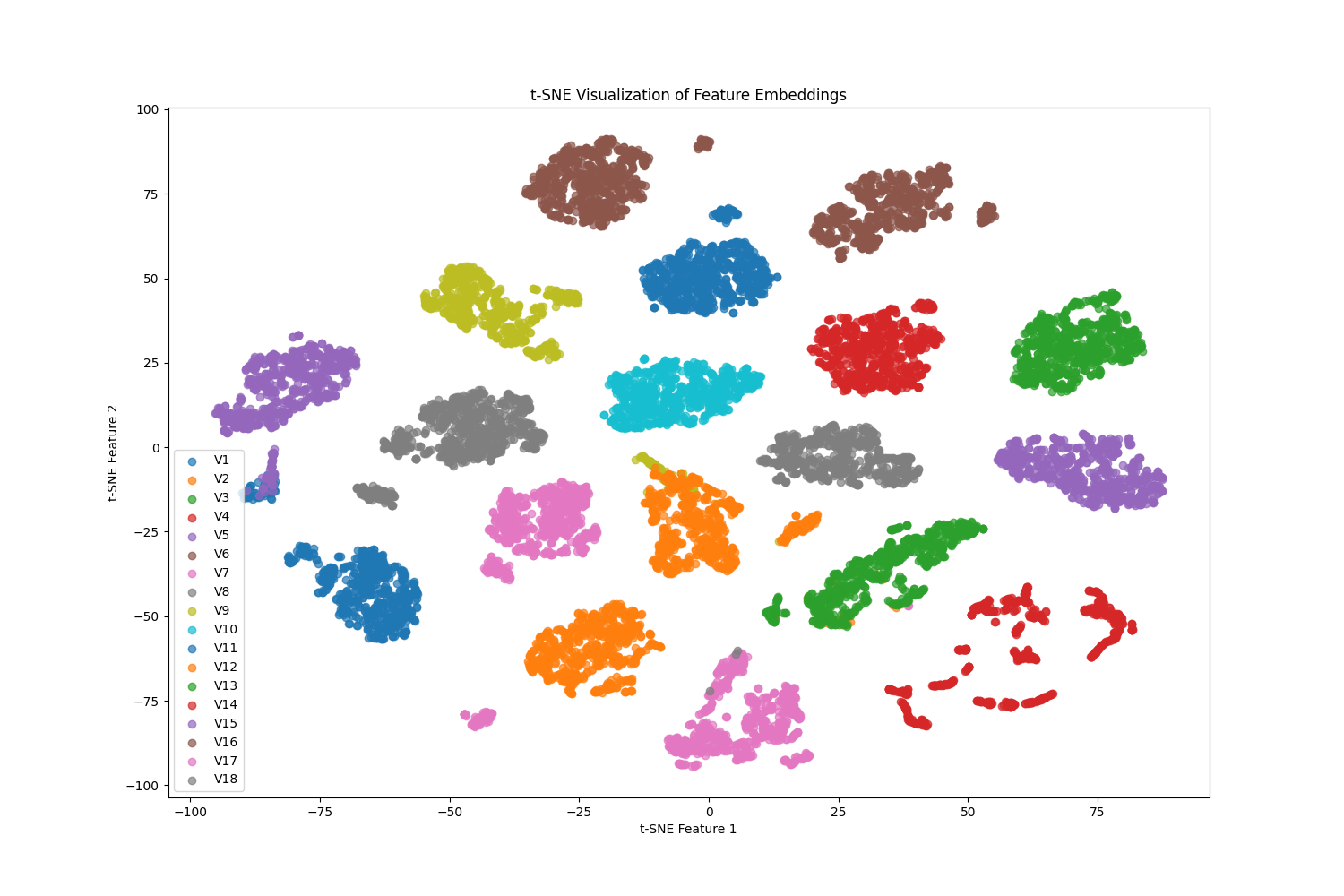}%
\label{fig:snr2034}}

\caption{The top layer represents the visualization of the original time-frequency image via t-SNE, while the bottom layer depicts the t-SNE visualization post-model classification. The sequence from left to right corresponds to an SNR ranging from Noiseless to 40dB.}
\label{fig11}
\end{figure*}

\subsection{Robustness to noise}
The performance of ST-GSResNet was evaluated across varying signal-to-noise ratios (SNR), namely at 20 dB, 30 dB, and 40 dB. The dataset comprised 18,000 samples, with each PQDs category comprising 1,000 samples. Seven hundred samples were employed for model training, while the remaining 300 samples were allocated for testing. Table \ref{tab:table4} illustrates that the classification accuracy decreases as signal-to-noise ratios (SNRs) decrease. At 20 dB, assigning categories C14 and C17 results in a classification accuracy drops to 88.3\% and 85.3\%, respectively. However, ST-GSesNet achieved an exceptional overall accuracy of 96.8\%.

Additionally, as the signal-to-noise ratio increased to 30 dB, the accuracy rose to 98.6\%. The results indicate that the CNN model incorporates advanced details despite utilizing only the most basic S-Transform in dataset generation. Our approach, ST-GSesNet, aims to enhance the model's recognition ability and exhibits remarkable tolerance to environmental noise. This can be attributed to the application of grouped convolution and incorporating the SE module in the model design. Grouped convolution enhances the model's expressive capability by introducing additional nonlinear transformations. Simultaneously, the SE module dynamically learns the significance of each channel, improving the model's performance by prioritizing essential feature channels and reducing the impact mitigating the influence of irrelevant information.

Following this, the confusion matrices originating from Table 4 are presented in Fig. \ref{fig:gsresnet}, which illustrate the accuracy of our proposed method, particularly against label noises. To highlight the robustness of the feature extraction capability inherent in ST-GSResNet, we visualize the feature distributions of the test samples via t-SNE, as depicted in Fig. \ref{fig11}. As can be seen from the top plot of the comparison in Fig. \ref{fig11}, the classification boundary becomes indistinguishable when time–frequency images, transformed by ST, are compressed to a 2D representation. However, as shown on the bottom side of Fig \ref{fig11}, after undergoing training through GSResNet, the features in the final layer become clearly defined. Even at an SNRdB level of 20, the boundaries between categories remain clearly distinguishable. This demonstrates ST-GSResNet's capacity to extract features that significantly contribute to the excellent classification of power quality disturbance signals.

\begin{table}[!t]
\caption{THE CLASSIFICATION ACCURACIES OBTAINED UNDER DIFFERENT NOISY ENVIRONMENTS\label{tab:table4}}
\centering

\begin{tabular}{c c c c c}
\toprule[1pt] 
Classes   &  Signal-to-Noise Ratio  &   &  &  \\
  &    20dB &  30dB & 40dB & No noise \\
\midrule
V1	& 98.0\%	& 96.7\%	& 100\%	& 100\%	 \\
V2	& 95.0\%	& 98.3\%	& 100\%	& 100\%	 \\
V3	& 99.7\%	& 100\%	& 100\%	& 100\%	 \\
V4	& 99.3\%	& 100\%	& 100\%	& 100\%	 \\
V5	& 99.3\%	& 100\%	& 100\%	& 100\%	 \\
V6	& 99.3\%	& 98.7\%	& 97.7\%	& 99.0\%	 \\
V7	& 93.3\%	& 99.7\%	& 100\%	& 100\%	 \\
V8	& 100\%	& 100\%	& 100\%	& 100\%	 \\
V9	& 100\%	& 100\%	& 100\%	& 100\%	 \\
V10	& 97.3\%	& 92.7\%	& 91.3\%	& 100\%	 \\
V11	& 98.3\%	& 100\%	& 100\%	& 100\%	 \\
V12	& 99.7\%	& 100\%	& 100\%	& 100\%	\\
V13	& 95.0\%	& 99.7\%	& 100\%	& 100\%	\\
V14	& 88.3\%	& 96.7\%	& 99.7\%	& 100\%	\\
V15	& 99.3\%	& 99.3\%	& 100\%	& 100\%	 \\
V16	& 93.3\%	& 100\%	& 100\%	& 100\%	 \\
V17	& 85.3\%	& 96.3\%	& 98.3\%	& 99.0\%	 \\
V18	& 94.0\%	& 99.3\%	& 99.7\%	& 99.3\%	 \\
Overall	& 96.5\%	& 98.8\%	& 99.3\%	& 99.9\%	 \\
\bottomrule[1pt] 
\end{tabular}
\end{table}

\subsection{Comparison with existing methods}

\begin{table}[!t]
\caption{COMPARISON WITH OTHER EXISTING METHODS\label{tab:table5}}
\centering
\begin{tabular}{p{3cm} p{0.8cm} p{0.8cm} p{0.3cm} p{0.3cm} p{0.3cm} }
\toprule[1pt] 
Method   &  PQD No.s &   Features No.s &  Accuracy(\%)  &  &  \\
&  & & 20dB & 30dB & 40dB \\

\midrule
CNN from Scratch \cite{salles2023}	& 6	& -	& 96.7	& 97.3 & - \\
ST and PNN \cite{wang2017}	& 9	& 4	& -	& 98.6	& 99.1 \\
ST and NSGA-II \cite{singh2018}	& 15	& 26	& 96.4	& 97.3	& 99.4 \\
DWT and PNN \cite{khokhar2017}	& 16	& 9	& 93.6	& 95.2	& 98.6 \\
HHT+WBELM \cite{sahani2018}	& 15	& 36	& 91.5	& -	& 95.6 \\
DBN+ELM \cite{swarnkar2023}	& 21	& 12	& 95.8	& 98.2	& 98.7 \\
SWT+EfficientNetB0 \cite{vishwanath2023} & 15 &  Auto &  92.8 & 99.0 & 99.2 \\
ST+ResNet50	& 18	& Auto	& 90.4	& 94.3	& 96.7 \\
ST+GSResNet	& 18	& Auto	& 96.5	& 98.8	& 99.3 \\

\bottomrule[1pt] 
\end{tabular}
\end{table}
This section presents a comparative analysis of the proposed method compared with other schemes in the field of PQDs detection and classification. Table \ref{tab:table5} summarizes the results of this comparative study, demonstrating that our proposed method surpasses other methods in both accuracy and the number of studied PQDs categories. Our proposed method utilizes the S-Transform (ST) for feature extraction and an enhanced ResNet-based technique for classification. A comparison of our proposed method with several recent studies reveals its superior accuracy and its ability to handle a greater number and complexity of PQDs categories. The method classifies signals with signal-to-noise ratios of 40 dB, 30 dB, and 20 dB, achieving accuracies of 99.3\%, 98.8\%, and 96.5\%, respectively. Overall, our proposed PQD detection and classification method surpasses other methods in terms of accuracy.

Prior investigations resulted in fewer classes of PQDs with a lower detection accuracy compared to our proposed method. Utkarsh Singh et al. \cite{wang2017} introduced an S-Transform (ST) and NSGA-II-based Randomized Binary Decision Tree (RBDT) classifier to classify 15 classes of PQDs. However, at 30 dB and 20 dB noise conditions, the accuracies are 97.3\% and 96.4\%, respectively \cite{singh2018}, which are lower than those achieved by our proposed method. In 2017, Khokhar, S. et al \cite{khokhar2017} . introduced a Discrete Wavelet Transform (DWT) and Probabilistic Neural Network-Artificial Bee Colony (PNN-ABC) based Randomized Binary Decision Tree (RBDT) classifier to classify 16 classes of PQDs. The accuracies were 98.6\%, 95.2\%, and 93.6\% for 40 dB, 30 dB, and 20 dB noise, respectively [21]. Compared to these two schemes, our approach demonstrates higher accuracy for 40 dB and 20 dB noise, incorporating more categories and more complex signal models using the same S-Transform-generated time-frequency maps. Sahani and Dash introduced a classifier based on Hilbert-Huang Transform (HHT) and Wavelet-Based Extreme Learning Machine (WBELM) for categorization with 40 dB and 20 dB noise. The accuracies were 95.6\% and 91.5\%, respectively \cite{sahani2018}. Swarnkar et al. designed a multivariate PQ interference identification algorithm using a mixture of S-Transform (ST), Hilbert Transform (HT), and Randomized Binary Decision Tree (RBDT). The accuracies were 98.7\%, 98.2\%, and 95.8\% for 40 dB, 30 dB, and 20 dB noise, respectively \cite{swarnkar2023}. These two schemes exhibit lower accuracy compared to our proposed method. Y. S. Upendra Vishwanath proposed a PQD classification scheme that combines synchrosqueezed wavelet transform (SWT) and EfficientNetB0 \cite{vishwanath2023}. We selected the results of the dataset from their paper, which also incorporates White Noise, for comparison. From the comparative analysis, it was observed that although our scheme's performance closely mirrors that of Vishwanath's scheme in the presence of 30dB and 40dB of noise, our scheme's performance improves by nearly 4 percentage points in the presence of 20dB noise. This indicates that our scheme exhibits a stronger resistance to noise.

Our proposed method outperforms all other methods, even when tested in a noisy environment and dealing with more complex PQDs classes. Overall, these results demonstrate the high effectiveness of our proposed method, surpassing other existing PQDs detection and classification methods.

\section{Conclusion}
In this study, we introduce the ST-GSResNet scheme for PQDs recognition and classification, with a focus on evaluating the model's performance. The experimental dataset comprises time-frequency images based on the S-Transform. Employing an improved ResNet model, we utilize grouped convolution to reduce the number of parameters, model complexity, and computational cost. Additionally, grouped convolution introduces additional nonlinear transformations, enriching feature representations to improve model performance and generalization. 

Furthermore, we introduce the SE module innovatively to enhance learning and focus on crucial parts of the input data, thereby improving recognition robustness and noise resistance. Compared to alternative deep learning methods, our approach demonstrates advantages such as reduced training time, increased accuracy, and fewer parameters. Future efforts will focus on gathering PQD datasets from various domestic and international regions for comprehensive classification and recognition assessments, further validating the method's practicality and effectiveness. Secondly, we employ knowledge distillation to reduce the model's size further, enabling its operation on smaller embedded devices to meet portability requirements in practical settings. 

With these enhancements and future investigations, the ST-GSResNet method will achieve more substantial progress in electrical energy and deep learning. We look forward to further exploring and applying the method to provide more reliable and efficient solutions for PQDs identification and classification in power systems.

\end{document}